\documentclass[prd,aps,nofootinbib]{revtex4}
\usepackage{framed}
\usepackage{bbold}
\usepackage{slashed}
\usepackage{graphicx,hyperref}
\usepackage{amsmath}
\usepackage{amssymb}
\usepackage{epsfig}
\usepackage{hhline}
\usepackage{soul}
\usepackage{tabularx,pifont,multirow}
\usepackage{enumitem}
\usepackage{colordvi}
\usepackage{soul}
\usepackage{xcolor}
\usepackage{yfonts}

\newcommand{\be}{\begin{equation}}
\newcommand{\ee}{\end{equation}}

\newcommand{\CC}{\Lambda}

\newcommand{\MPl}{M_{\rm Pl}}
\newcommand{\gCS}{g$\mathcal C\mathcal S\ $}
\newcommand{\CS}{$\mathcal C\mathcal S$}

\definecolor{darkgreen}{rgb}{0,0.3,0.05}
\makeatletter                                                         %
\newcommand*\rel@kern[1]{\kern#1\dimexpr\macc@kerna}                  %
\newcommand*\widebar[1]{                                              %
  \begingroup                                                         %
  \def\mathaccent##1##2{                                              %
    \rel@kern{0.8}                                                    %
    \overline{\rel@kern{-0.8}\macc@nucleus\rel@kern{0.2}}             %
    \rel@kern{-0.2}                                                   %
  }                                                                   %
  \macc@depth\@ne                                                     %
  \let\math@bgroup\@empty \let\math@egroup\macc@set@skewchar          %
  \mathsurround\z@ \frozen@everymath{\mathgroup\macc@group\relax}     %
  \macc@set@skewchar\relax                                            %
  \let\mathaccentV\macc@nested@a                                      %
  \macc@nested@a\relax111{#1}                                         %
  \endgroup                                                           %
}                                                                     %
\makeatother                                                          %

\begin{document}

\preprint[\leftline{KCL-PH-TH/2019-{\bf 89}}

%

\title{\Large {\bf  Quantum Anomalies in String-Inspired Running
Vacuum Universe: Inflation and Axion Dark Matter} \vspace{0.0cm}}

\author{\large \bf Spyros Basilakos$^{a,b}$, Nick E. Mavromatos$^{c}$ and Joan Sol\`a Peracaula$^d$ \vspace{0.5cm}}

\affiliation{$^a$Academy of Athens, Research Center for Astronomy and Applied Mathematics, Soranou Efessiou 4, 115 27 Athens, Greece. \\
$^b$ National Observatory of Athens, Lofos Nymfon,
11852, Athens, Greece. \vspace{0.5cm}\\
$^c$Theoretical Particle Physics and Cosmology Group, Physics Department, King's College London, Strand, London WC2R 2LS.
\vspace{0.5cm}\\
 $^{d}$Departament de F\'\i sica Qu\`antica i Astrof\'\i sica, \\ and \\ Institute of Cosmos Sciences (ICCUB), Universitat de Barcelona, Avinguda Diagonal 647 E-08028 Barcelona, Catalonia, Spain.}


\begin{abstract}
\vspace{0.05cm}
In this letter, we elaborate further on a Cosmological ``Running-Vacuum'' type model for the Universe, suggested previously by the authors~\cite{bmas,bmasPRD},
within the context of a string-inspired effective theory in the presence of a Kalb-Ramond (KR) gravitational axion field which descends from the antisymmetric tensor of the massless gravitational string multiplet. In the presence of this field, which has anomalous CP violating interactions with the gravitons, primordial gravitational waves induce gravitational anomalies, which in turn are responsible for the appearance of $H^2$ and $H^4$ contributions to the vacuum  energy density, these terms being  characteristic of generic  ``running-vacuum-model (RVM) type'', where $H$ is the Hubble parameter. In this work we prove in detail the appearance of the $H^4$ terms due to gravitational-anomaly-induced  condensates in the energy density of the primordial Universe, which can self-consistently induce inflation, and subsequent exit from it, according to the generic features of RVM. {We also argue in favour of the robustness of our results, which were derived within an effective low-energy field theory approach, against Ultra Violet
completion of the theory}. During the radiation and matter-dominated eras, gravitational anomalies cancel, as required for the consistency of the quantum matter/radiation field theory. However, chiral and QCD-axion-type anomalies survive and have important consequences for both cosmic magnetogenesis and axionic dark matter in the Universe. {Finally, the stringy RVM scenario presented here predicts quintessence-like dynamical dark energy for the current Universe, which is compatible with the existing fitting analyses of such model against observations.}
\end{abstract}
\maketitle

{ \it Introduction}

\vspace{0.5cm}
In spite of the very good agreement of the $\CC$CDM model ({the current standard or ``concordance'' model of cosmology}) with the currently available plethora of observational data~\cite{Planck}, nonetheless the model appears to be currently in tension with some important measurements~\cite{s8},
associated with the value of $\sigma_8$ {(the mean matter fluctuations in spheres of {radius} $8h^{-1}$~Mpc, {with $h$ the reduced Hubble constant})  and the disparate current Hubble parameter $H_0$ values obtained independently from {measurements} of the local and the early universe}.  Whether these tensions are the result of yet unknown systematic errors or hint some underlying new Physics is still unclear.  Therefore, there remains strong the possibility that a deviation from the $\CC$CDM model could provide an explanation for such discrepancies{\,\cite{R19}}.

Among the several existing candidates beyond the $\CC$CDM, which can alleviate these tensions, here we
focus on the ``running vacuum model'' (RVM)~\cite{ShapSol,Fossil07}.  {For a review see\, \cite{RVMreviews}}.  The detailed phenomenology of this framework, and its advantages as compared to
$\CC$CDM in fitting the
current data, has been amply discussed in several works~\cite{rvmpheno},  {including its scalar field description\,\cite{JCAP2019}. It is also remarkable that frameworks mimicking the RVM (even beyond the GR paradigm) may acquire the ability to alleviate those tensions\,\cite{BD2019}. This fact is actually very important since it is a solid motivation for us to explore fundamental models which can lead to such phenomenological success thanks precisely to their ``effective RVM behavior''.}

{Attempts to link the evolution of the vacuum energy density with a fundamental framework of  Quantum Field Theory in curved spacetime can be found in the above mentioned RVM papers} and, in the case of Supergravity inflationary models, in \cite{bamasol}. Recently~\cite{bmas}, we have also presented a potential connection of the  RVM with string effective actions, through gravitational anomalies, that exist due to gravitational-wave perturbations in the early Universe --- {see \cite{bmasPRD} for a comprehensive exposition}. Specifically, we studied a four-dimensional string-inspired  version of the RVM, based on the  low-energy effective action of graviton and antisymmetric tensor  fields of the massless (bosonic) string gravitational multiplet. The latter, provides the minimal field content of the inflationary universe in these kind of theories, and is responsible for the effective RVM behavior of the vacuum energy density through $\sim H^4$ contributions. {It has been shown that, once such high power of $H$ has been generated, the inflationary regime with graceful exit is warranted~\cite{rvmInflation} and the cosmic evolution satisfies the Generalized Second Law of Thermodynamics\,\cite{Yu2019}}. Remarkably, condensates of graviton fluctuations have been argued to provide dynamically such $\sim H^4$ contributions, which are responsible for the early de Sitter phase~\cite{rvmInflation}. Furthermore, it has recently been found in \cite{bmas,bmasPRD}, that, at the exit from the inflationary phase,  there exist undiluted backgrounds of the antisymmetric tensor field which violate spontaneously the Lorentz and CPT symmetry, thus providing
the interesting possibility of baryogenesis through leptogenesis~\cite{bms}. Moreover, the generation of  chiral matter at the exit from inflation was held responsible for the cancellation of gravitational anomalies, but chiral anomalies may remain uncompensated, and have been considered responsible for cosmic magnetogenesis.

The purpose of the present Letter is first to scrutinise the r\^ole of the gravitational anomalies in the early Universe in
producing dynamical inflation, without the need of fundamental inflatons, {since the RVM already contains an alternative ingredient for inflation, which is the aforementioned $\sim H^4$ term}.  Second, we argue in favour of the robustness of the results of ref.~\cite{bmas,bmasPRD}, against Ultraviolet Completion (UV) of the effective theory. Finally, we discuss the physics of the post inflationary Universe in more detail, by arguing that QCD-axion-type anomalies, that may survive during the QCD epoch, have important consequences for axionic dark matter,  the source of which can be the KR field itself. Indeed, we show that
non-perturbative instanton effects can generate a potential (and hence a mass) for the KR axion, thus implying its r\^ole as conventional Cold Dark Matter, consistently with phenomenology. Moreover, the presence of the KR axion triggers the more moderate effective  $\sim H^2$ behavior of the vacuum density at late epochs, {which can be responsible for a dynamical form of dark energy at present and hence be susceptible of detection\,\cite{rvmpheno}.}

\vspace{0.5cm}
{\it Gravitational Anomalies and String Effective Actions for Running Vacuum.}
\vspace{0.5cm}

Our starting framework is the four-dimensional string-inspired  version of the RVM, based on critical-string low-energy effective actions of the graviton and antisymmetric tensor (spin-one) Kalb-Ramond (KR) fields of the massless (bosonic) string gravitational multiplet~\cite{gsw,string,kaloper}:

\begin{align}\label{sea2}
S_B =-&\; \int d^{4}x\sqrt{-g}\Big( \dfrac{1}{2\kappa^{2}}\, R + \frac{1}{6}\, {\mathcal H}_{\lambda\mu\nu}\, {\mathcal H}^{\lambda\mu\nu} + \dots \Big),
\end{align}
where
${\mathcal H}_{\mu\nu\rho} \equiv \kappa^{-1} \partial_{[\mu}\, B_{\nu\rho]}$ is the field strength of the KR field $B_{\mu\nu}$, with $\kappa=\MPl^{-1}$, and  $\MPl = 2.4 \times 10^{18}$~GeV  is the four-dimensional reduced Planck mass; the symbol $[\dots ]$ denotes complete antisymmetrisation of the respective indices.  {The dots $\dots$  in Eq.\,\eqref{sea2}  involve, on the one hand, the higher order terms in  the expansion of the effective action in powers of $\alpha^\prime = M_s^{-2}$  (the Regge slope of the string), with $M_s$ the string mass scale (which is, in general, different from $M_{\rm Pl}$~\cite{string}); and, on the other hand,  a (non-perturbatively generated) effective potential for the dilaton. More details are given in the Appendix.}

{An important remark is due at this point. Although the model \eqref{sea2}, and its Bianchi-anomaly constraint \eqref{modbianchi2}, to be discussed below, are inspired by string theory considerations, the theory might be considered independent of strings, as a self consistent field theory of graviton and KR degrees of freedom. This is the point of view we take in this article. Nonetheless, as we explain in the Appendix, our main conclusions remain unaffected when features from microscopic string theory are taken into account, such as, for instance, the inclusion of the dilaton degree of freedom, existing in string models, through its equations of motion, providing extra constraints, as well as the r\^ole of higher order stringy-inspired terms in \eqref{sea2}. As we show in  the Appendix, the solution of a constant dilaton, adopted in our work and in \cite{bmasPRD}, appears self consistently in that framework.}

The KR-field-strength terms ${\mathcal H}^2$ in (\ref{sea2}) can be absorbed (up to an irrelevant total divergence) into a {\it contorted} generalised curvature~\cite{gsw,string}
$\overline R (\overline \Gamma)$, with a ``torsional connection''~\cite{hehl} $\overline \Gamma$, corresponding to a contorsion tensor proportional to ${\mathcal H}_{\mu\nu}^\rho$ field strength,
\begin{align}\label{torcon}
{\overline \Gamma}_{\mu\nu}^{\rho} = \Gamma_{\mu\nu}^\rho + \frac{\kappa}{\sqrt{3}}\, {\mathcal H}_{\mu\nu}^\rho  \ne {\overline \Gamma}_{\nu\mu}^{\rho}~,
\end{align}
where $\Gamma_{\mu\nu}^\rho = \Gamma_{\nu\mu}^\rho$ is the torsion-free Christoffel symbol.\footnote{Exploiting local field redefinition ambiguities~\cite{string,kaloper}, which do not affect the perturbative scattering amplitudes, one may extend the above conclusion to the quartic order in derivatives, that is,
to the ${\mathcal O}(\alpha^\prime)$ effective low-energy action, which includes Gauss-Bonnet quadratic curvature invariants. In terms of a generalised curvature, the Gauss-Bonnet-type invariants are not total derivatives, and simply correspond to higher than quadratic order $\mathcal H_{\mu\nu\rho}$ terms.}
The torsion interpretation of $\mathcal H_{\mu\nu\rho}$ is of crucial importance when one discusses effective actions in the presence of fermions~\cite{bmas,bmasPRD}, as we shall discuss later on.

Due to its definition as a curl of the spin-one antisymmetric tensor $B_{\mu\nu}$,
the 3-form ${{\mathcal H}}_{\mu\nu\rho}$ satisfies the following Bianchi identity {at the classical level}
\begin{equation}\label{bianchi}
\partial_{[\mu}\, {{\mathcal H}}_{\nu\rho\sigma]} = 0,
\end{equation}
by construction. However, in string theory, in the presence of gauge and gravitational fields, cancellation of anomalies {at the quantum level} requires the modification of the definition of the KR field strength $\mathcal H_{\mu\nu\rho}$
by appropriate gauge (Yang-Mills (Y)) and Lorentz (L) Chern--Simons three-forms~\cite{gsw}

The modified Bianchi identity constraint  can be expressed in the usual tensor notation as follows:
\begin{align}\label{modbianchi2}
& \varepsilon_{abc}^{\;\;\;\;\;\mu}\, {\mathcal H}^{abc}_{\;\;\;\;\;\; ;\mu}
 =  \frac{\alpha^\prime}{32\, \kappa} \, \sqrt{-g}\, \Big(R_{\mu\nu\rho\sigma}\, \widetilde R^{\mu\nu\rho\sigma} -
F_{\mu\nu}\, \widetilde F^{\mu\nu}\Big) \equiv \sqrt{-g}\, {\mathcal G}(\omega, \mathbf{A}),
\end{align}
where the semicolon denotes covariant derivative with respect to the standard
Christoffel connection, and
the dual $\widetilde \dots$ of the gauge field strength is defined  as: $\widetilde F_{\mu\nu} = \frac{1}{2} \varepsilon_{\mu\nu\rho\sigma}\, F^{\rho\sigma}$. The term
$\sqrt{-g}\, \Big( R_{\mu\nu\rho\sigma}\, \widetilde R^{\mu\nu\rho\sigma} - F_{\mu\nu}\, \widetilde F^{\mu\nu} \Big)$ in (\ref{sea3}) is the  \emph{Hirzebruch-Pontryagin topological density}, also known as the Chern-Simons (\CS) topological, or {\it mixed anomaly}, term~\cite{jackiw}. It contains a gravitational (henceforth referred to as the \gCS-term) and an ordinary gauge part, both being individually total derivatives:
\begin{align}\label{pontryaginA}
&\sqrt{-g} \, \Big(R_{\mu\nu\rho\sigma}\, \widetilde R^{\mu\nu\rho\sigma} - F_{\mu\nu}\, \widetilde F^{\mu\nu} \Big) = \sqrt{-g} \, {\mathcal K}_{\rm mixed}^\mu (\omega)_{;\mu} = \partial_\mu \Big(\sqrt{-g} \, {\mathcal K}_{\rm mixed}^\mu (\omega) \Big)\nonumber \\ &
= 2 \, \partial_\mu \Big[\epsilon^{\mu\nu\alpha\beta}\, \omega_\nu^{ab}\, \Big(\partial_\alpha \, \omega_{\beta ab} + \frac{2}{3}\, \omega_{\alpha a}^{\,\,\,\,\,\,\,c}\, \omega_{\beta cb}\Big)  - 2 \epsilon^{\mu\nu\alpha\beta}\, \Big(A^i_\nu\, \partial_\alpha A_\beta^i + \frac{2}{3} \, f^{ijk} \, A_\nu^i\, A_\alpha^j \, A_\beta^k \Big)\Big],
\end{align}
with Latin letters $i,j,k$ being gauge group indices, and $\sqrt{-g}\, {\mathcal K}_{\rm mixed}^\mu$ denoting the mixed-anomaly current density. {Here $\omega_\mu^{ab}$ is the spin-connection one-form and $A_\mu^i$ are the ordinary gauge fields, labeled respectively $\omega$ and $\mathbf{A}$ for short.}
For more technical details, see \cite{bmasPRD}.

Since the anomaly ${\mathcal G}(\omega, \mathbf{A})$ is an exact one loop result, one may implement the Bianchi identity (\ref{modbianchi2}) as a $\delta$-functional constraint in the quantum path integral of the action (\ref{sea2}) over the fields ${\mathcal H}$, $\mathbf{A}$, and $g_{\mu\nu}$, and express the latter in terms of a Lagrange multiplier (pseudoscalar) field~\cite{kaloper} $b(x)/\sqrt{3}$ (where the normalisation factor $\sqrt{3}$ is inserted so that the field $b(x)$ will acquire a canonical kinetic term).
Inserting such a constraint
into the path integral with respect to the action (\ref{sea2}), and integrating over the ${\mathcal H}$ field, one obtains an effective action in terms of the anomaly and a
canonically normalised dynamical, {\it massless}, KR axion field $b(x)$~\cite{kaloper,bmas,bmasPRD}
\begin{align}\label{sea3}
S^{\rm eff}_B =&\; \int d^{4}x\sqrt{-g}\Big[ -\dfrac{1}{2\kappa^{2}}\, R + \frac{1}{2}\, \partial_\mu b \, \partial^\mu b   + \sqrt{\frac{2}{3}} \, \frac{\alpha^\prime}{96\, \kappa} \, b(x) \, \Big(R_{\mu\nu\rho\sigma}\, \widetilde R^{\mu\nu\rho\sigma} - F_{\mu\nu}\, \widetilde F^{\mu\nu}\Big) + \dots \Big],
\end{align}
where the dots $\dots$ denote gauge, as well as higher derivative, terms appearing in the string effective action, that we ignore for our discussion here.  {The reader should notice in this respect that, in view of \eqref{pontryaginA}, the anomaly terms in \eqref{sea3} are quadratic in derivatives}.

We thus observe that, in view of the anomaly, the KR axion field {in \eqref{sea3} couples to both gravitational (g$\mathcal C\mathcal S$) and gauge-field \CS\, terms (cf. \eqref{pontryaginA}). These interactions are P and T violating, and hence in view of the overall CPT invariance of the quantum theory \eqref{sea3}, also CP violating.\footnote{In this respect, the reader should recall that any breaking of the Lorentz and CPT symmetry by the KR axion
background, which we considered in our analysis so far~\cite{bmas,bmasPRD}, and shall employ in this work, implies a {\it spontaneous} violation of these symmetries.} They play quite an  important r\^ole for our purposes in this work and in \cite{bmas}, {see also  \cite{bmasPRD} for an expanded exposition}.

\vspace{0.5cm}

{\it Primordial Gravitational Waves and Anomalies.}

\vspace{0.5cm}

In the early Universe, before and during inflation, we assume that only fields from the gravitational multiplet of the string exist, which implies that our effective action pertinent to the dynamics of the inflationary period, is given by (\ref{sea3}) upon setting the gauge fields to zero, $\mathbf{A}=0$. Thus, to describe the dynamics of the beginning and the inflationary period of the Universe, we use the following effective action
involving only the KR axion and the gravitational field:
\begin{align}\label{sea4}
S^{\rm eff}_B =&\; \int d^{4}x\sqrt{-g}\Big[ -\dfrac{1}{2\kappa^{2}}\, R + \frac{1}{2}\, \partial_\mu b \, \partial^\mu b
+   \sqrt{\frac{2}{3}}\,
\frac{\alpha^\prime}{96 \, \kappa} \, b(x) \, R_{\mu\nu\rho\sigma}\, \widetilde R^{\mu\nu\rho\sigma} + \dots \Big] \nonumber \\
=&\; \int d^{4}x\, \sqrt{-g}\Big[ -\dfrac{1}{2\kappa^{2}}\, R + \frac{1}{2}\, \partial_\mu b \, \partial^\mu b  -
 \sqrt{\frac{2}{3}}\,
\frac{\alpha^\prime}{96 \, \kappa} \, \partial_\mu b(x) \, {\mathcal K}^\mu + \dots \Big]~,
\end{align}
where in the second equality we have partially integrated the CP violating anomaly term. {Notice that here the anomaly current ${\mathcal K}^\mu$ contains only the gravitational part of \eqref{pontryaginA}.}

For a complete study of the equation of state  {(EoS)} of the gravitational-KR-axion fluid we refer the reader to \cite{bmasPRD}. For our purposes
here we notice that, in the presence of the g$\mathcal C\mathcal S$ term,  the corresponding Einstein's equations derived {from variation of the action \eqref{sea4} (using the first expression in that equation on this occasion)} have the form
\begin{align}\label{einsteincs}
R^{\mu\nu} - \frac{1}{2}\, g^{\mu\nu} \, R - \sqrt{\frac{2}{3}}\,
\frac{\alpha^\prime\, \kappa}{12} \,{\mathcal C}^{\mu\nu} = \kappa^2 \, T^{\mu\nu}_{b},
\end{align}
where
\begin{align}\label{tbstress}
T_b^{\mu\nu} = \partial^\mu b\, \partial^\nu b -  \frac{1}{2}\, g^{\mu\nu}\,  \Big(\partial^\alpha b \, \partial_\alpha b\Big)~,
\end{align}
is the stress tensor of the massless KR axion, and
\begin{align}\label{cotton}
\mathcal C^{\mu\nu} &=  -\frac{1}{2} \Big[v_\sigma \, \Big( \varepsilon^{\sigma\mu\alpha\beta} R^\nu_{\, \, \beta;\alpha}  +
\varepsilon^{\sigma\nu\alpha\beta} R^\mu_{\, \, \beta;\alpha}\Big)   + v_{\sigma\tau} \, \Big(\widetilde R^{\tau\mu\sigma\nu} +
\widetilde R^{\tau\nu\sigma\mu} \Big)\Big]   = - \frac{1}{2} \Big[\Big(v_\sigma \, \widetilde R^{\lambda\mu\sigma\nu}\Big)_{;\lambda}  + \, (\mu \leftrightarrow \nu)\Big]\, , \nonumber \\ v_{\sigma} &\equiv \partial_\sigma b = b_{;\sigma}, \,\,v_{\sigma\tau} \equiv  v_{\tau; \sigma} = b_{;\tau;\sigma},
\end{align}
is the Cotton tensor~\cite{jackiw},\footnote{We note, for completeness, that the (3+1)-dimensional Cotton tensor constructed in \cite{jackiw} and used here  is different from another four-space-time-dimensional tensor, also called (1+3)-dimensional Cotton, which was constructed in \cite{garcia}
as a direct extension of a three-dimensional Cotton tensor, associated with (1+2)-dimensional \gCS theories.} {arising from the variation of the \gCS -term in \eqref{sea4}} with respect to the gravitational field:
$\delta \Big[ \int d^4x \sqrt{-g} \, b \, R_{\mu\nu\rho\sigma}\, \widetilde R^{\mu\nu\rho\sigma} \Big] = 4 \int d^4x \sqrt{-g} \, {\mathcal C}^{\mu\nu}\, \delta g_{\mu\nu} = -
4 \int d^4x \sqrt{-g} \, {\mathcal C}_{\mu\nu}\, \delta g^{\mu\nu}$.
As follows from its definition \eqref{cotton}, and properties of the Riemann tensor, the Cotton tensor is traceless~\cite{jackiw}
\begin{align}\label{tracecot}
g_{\mu\nu}\, \mathcal C^{\mu\nu}= 0~.
\end{align}
In standard situations, general coordinate diffeomorphism invariance, would imply the conservation of the matter stress tensor,
$T^{\mu\nu}_{b \,\,\,\,; \nu}=0$. Because of the curvature tensor Bianchi identity, the Einstein tensor
 $R^{\mu\nu} - \frac{1}{2}\, g^{\mu\nu} \, R $,
also obeys such a covariant  conservation law, but this is {\it not} the case for the Cotton-tensor, as one can readily check from \eqref{cotton}~\cite{jackiw}. {As a consequence, by taking the covariant derivative on both sides of \eqref{einsteincs} we find}
\begin{align}\label{csder}
\sqrt{\frac{2}{3}}\,\frac{\alpha^\prime\, \kappa}{12} \, {\mathcal C}^{\mu\nu}_{\,\,\,\,\,\,\,;\mu} = -\sqrt{\frac{2}{3}}\,\frac{\alpha^\prime\, \kappa}{12} \, \frac{1}{8} (\partial^\nu b) \, R^{\alpha\beta\gamma\delta} \, \widetilde R_{\alpha\beta\gamma\delta} = - \kappa^2 \, T^{\mu\nu}_{b\,\,\,\,\,\,\,;\mu}~.
\end{align}

Thus, in the presence of {\it gravitational} anomalies, the diffeomorphism invariance, and hence the conservation of $T^{\mu\nu}_b$ appears to be in trouble, unless one deals with specific gravitational backgrounds, as the ones pertaining to the FLRW Universe of interest to us here, for which the Pontryagin density {\it vanishes} identically  $R_{\mu\nu\rho\sigma} \widetilde R^{\mu\nu\rho\sigma} = 0$.
Nonetheless, there is no fundamental issue here. Indeed, notice, from \eqref{csder}, that there is a {\it conserved}  modified stress-energy tensor
\begin{align}\label{cons}
\kappa^2 \, {\widetilde T}_{b + g\mathcal C\mathcal S}^{\mu\nu} \equiv \sqrt{\frac{2}{3}}\,\frac{\alpha^\prime\, \kappa}{12} \mathcal C^{\mu\nu} + \kappa^2 T_b^{\mu\nu} \quad \Rightarrow \quad  {\widetilde T}_{b + g\mathcal C\mathcal S \,; \mu}^{\mu\nu} =0~,
\end{align}
and hence, the non-vanishing divergence of the Cotton tensor in anomalous backgrounds simply expresses the non-trivial interactions between the axion $b$-field and gravity, leading to energy exchange\,\footnote{A similar situation characterises the interactions between dilatons and Gauss-Bonnet quadratic curvature terms in $\mathcal O (\alpha^\prime)$ string effective actions~\cite{kanti}.
That there is no fundamental problem also follows in our case by the fact that, both, the effective action \eqref{sea4}, and the underlying microscopic string theory, which leads to it at low energies, are fully covariant.}

As we shall discuss later on, the contributions of the (quadratic in Riemann curvature tensor) g$\mathcal C\mathcal S$ term to the time component of the modified stress-energy tensor $\widetilde T_{00}$ ({\it i.e.} the energy density) in our mean-field ground state solution will turn out to be {\it negative}, in a similar spirit to the energy contributions of the dilaton-Gauss-Bonnet term in
${\mathcal O}(\alpha^\prime)$ string effective actions~\cite{kanti}.

In a de Sitter-type cosmological  background, characterised by an (approximately)  constant Hubble parameter $H \simeq$ const., the average $\langle \dots \rangle$ of the anomalous g$\mathcal C\mathcal S$ term {(over quantum fluctuations about the  de Sitter cosmological background)} of the metric tensor of {\it primordial gravitational-wave} type)  yields a non-zero result~\cite{stephon}:
\begin{align}\label{rrt}
  \langle R_{\mu\nu\rho\sigma}\, \widetilde R^{\mu\nu\rho\sigma} \rangle  &= \frac{16}{a^4} \, \kappa^2\int_0^\mu \frac{ 4\pi \, k^2\, d k}{(2\pi)^3} \, \frac{H^2}{2\, k^3} \, k^4 \, \Theta + {\rm O}(\Theta^3) , \nonumber \\&
 \Theta {\equiv }\sqrt{\frac{2}{3}}\, \frac{\alpha^\prime \, \kappa}{12} \, H \,  {\dot {\overline b}} \, \ll 1,
  \end{align}
under the
slow-roll assumption for the KR axion field $b(t)$,
 \begin{equation}\label{slowroll}
{\dot {\overline b}} \ll  H/\kappa.
\end{equation}
Here and in what follows, the notation $\overline b(t)$ indicates a background solution of the equations of motion for the KR field ({\it cf.}
Eqs.~\eqref{krbeom},  \eqref{krbeom2} below). The overdot denotes derivative with respect to the cosmic time.
The $\langle \dots \rangle$ is
 calculated in~\cite{stephon}, using appropriate Green functions to leading order in $k \, \eta \gg 1$, where $k$ is the standard Fourier scale variable for the gravitational wave graviton modes,
and $\eta$ is the conformal time $d\eta =  \frac{dt}{a(t)} \,  \Rightarrow \, \eta = \frac{1}{H}\, \exp (-Ht)$, during inflation, which runs in the opposite direction of the cosmic time $t$.
{Because the integral in \eqref{rrt} is quartically divergent, we use an ultraviolet cutoff $\mu$ for $k$. Following ~\cite{stephon} we take the range}
\begin{equation}\label{cutoff}
0 < k\,  < \mu /(\eta H).
\end{equation}

The smallness of $\Theta$ {(i.e. $|\Theta|\ll 1$)}, when combined with \eqref{slowroll}, implies the {\it sufficient} condition
\begin{align}\label{HMs}
{H^2/M^2_s \ll 12\sqrt{3/2}\ \Rightarrow\ \ \ H/M_s \ll 3.83\,.}
\end{align}
We take  for concreteness the inflationary Hubble parameter $H$ in the range
\begin{equation}\label{Hinfl}
\frac{H}{\MPl}\sim  10^{-4} ~,
\end{equation}
as implied by the cosmological data~\cite{Planck}. From \eqref{rrt} and \eqref{Hinfl}, then, we arrive at the {\it sufficient} condition
\begin{align}
\frac{M_{\rm Pl}}{M_s} \ll 3.83 \times (10^4 - 10^5).
\end{align}
We shall make use of this result later on.

From (\ref{sea4}) -- {using this time the expression on the second line} -- it follows that the classical equations of motion of the KR axion field $b(x)$, imply the existence of
backgrounds $\overline b$ that satisfy
\begin{align}\label{krbeom}
\partial_{\alpha}\Big[\sqrt{-g}\Big(\partial^{\alpha}\bar{b}  -  \sqrt{\frac{2}{3}}\,
\frac{\alpha^\prime}{96 \, \kappa} \, {\mathcal K}^{\alpha}  \Big)\Big] = 0,
\end{align}
which, on the assumption of homegeneity and isotropy of the inflationary space time, would imply a partial solution~\cite{bmas,bmasPRD}:
\begin{align}\label{krbeom2}
\dot{\overline{b}}  =  \sqrt{\frac{2}{3}}\, \frac{\alpha^\prime}{96 \, \kappa} \, {\mathcal K}^{0}.
\end{align}
Eq.~\eqref{krbeom2} is a mathematically consistent relation, since both $\partial_\mu b$ and ${\mathcal K}_\mu$ are (covariant) {\it axial} four-vectors. This solution implies a background for the KR axion field that breaks, {\it spontaneously} Lorentz, CP and CPT symmetry,  which is crucial for leptogenesis in the post inflationary period~\cite{bmas,bmasPRD}, following the mechanism of \cite{bms}.
In fact the {\it masslessness} of the KR axion $b$ can be understood by viewing this pseudoscalar field as the {\it Goldstone-Boson} of the spontaneously broken Lorentz symmetry~\cite{aben}.

From the anomaly equation (\ref{pontryaginA}), and taking into account that
 the gravitational waves are weak perturbations, since their intensity tends to be proportional to $|\Theta| \ll 1$~\cite{stephon},
 one obtains~\cite{bmas,bmasPRD}
\begin{align}\label{k01}
& \frac{d}{dt}  \Big(\sqrt{-g}\, {\mathcal K}^0 (t) \Big) = \langle \sqrt{-g} \, R_{\mu\nu\rho\sigma}\, \widetilde R^{\mu\nu\rho\sigma} \rangle
 \simeq \sqrt{-g} \, \langle R_{\mu\nu\rho\sigma}\, \widetilde R^{\mu\nu\rho\sigma} \rangle \simeq \sqrt{-g} \, \frac{1}{\pi^2} \Big(\frac{H}{M_{\rm Pl}}\Big)^2 \, \mu^4\, \Theta  \nonumber \\& \simeq
 \Big[{\frac{1}{3\pi^2\times 6\times 96}}\,  \Big(\frac{H}{M_{\rm Pl}}\Big)^3 \, \Big(\frac{\mu}{M_s}\Big)^4 \,  M_{\rm Pl}\Big]\, \times \, \Big(\sqrt{-g} \, {\mathcal K}^0 (t(\eta))\Big),
 \end{align}
where we used \eqref{rrt} and \eqref{krbeom2}.

Since, during inflation, $H$ remains approximately constant, (\ref{k01}) can be integrated over the inflationary period,
yielding
\begin{align}\label{k02}
{\mathcal K}^0 (t(\eta))  \simeq {\mathcal K}^0_{\rm begin} (t=0) \, \exp\Big[  - 3H\, t(\eta)\, \mathcal A\Big]~, \qquad
{\mathcal A} \equiv  1  -  {\frac{1}{3\pi^2\times 18\times 96} }\,  \Big(\frac{H}{M_{\rm Pl}}\Big)^2 \, \Big(\frac{\mu}{M_{s}}\Big)^4,
\end{align}
where we have set the beginning of inflation at $t=0$ ($\eta = H^{-1}$) and its end at $t \to +\infty$ ($\eta \to 0$), so that in conformal time units the duration of inflation is $|\Delta \eta| = 1/H$~\cite{stephon,bmas,bmasPRD}.
The value ${\mathcal K}^0_{\rm begin} (t(\eta=H^{-1})$, which on account of (\ref{krbeom2}) corresponds to an initial condition for the cosmic time derivative of the KR axion, $\dot{\overline{b}}(0)$, is a boundary condition to be determined phenomenologically, as we shall discuss later on.

In \cite{bmas,bmasPRD} it was observed that if the factor ${\mathcal A} \simeq 0$ then $\mathcal K^0$ is approximately constant, for
a momentum cutoff on the graviton modes of order\begin{align}\label{A=0}
{\mathcal A} \simeq 0 \quad \stackrel{({\it cf.} \eqref{k02})}{\Rightarrow}  \quad
\frac{\mu}{M_s} \simeq 15 \, \Big(\frac{\MPl}{H}\Big)^{1/2}~,
 \end{align}
 where the $\simeq$ in the above relations are to be interpreted as within an error of order of at most a $\%$.\footnote{Indeed, an approximately constant $\mathcal K^0$ in \eqref{k02} is guaranteed provided that at the end of the inflationary period its value is diminished no more than an order of magnitude, that is
 $$
 {\mathcal K}^0_{\rm end} (t_{\rm end}) \simeq \Big(e^{-1} - e^{-2}\Big)\, {\mathcal K}^0_{\rm begin} (t(\eta=H^{-1}))
 $$
 Taking in to account that, in units of cosmic Robertson-Walker time $t$, the end of inflation occurs for $H \, t_{\rm end} \sim {\mathcal N}$, with ${\mathcal N}$, the number of e-foldings, which is expected from the data~\cite{Planck} to be of order $\mathcal N= {\mathcal O}(60-70)$, we thus observe from {the above equation} that {the following condition}
 %
 $$
0 \lesssim  {\mathcal A} \lesssim \xi \, (3{\mathcal N})^{-1} \sim  \xi (0.0048 - 0.0056 ), \quad {\xi ={\cal O}(1)},
 $$
 suffices for our purposes, which leads to the aforementioned uncertainty of at most a $\%$ in the value of $\mu$ in \eqref{A=0},
 \begin{align*}\label{A=0N}
 \frac{\mu}{M_s} &\simeq 15 \Big( 1-\frac{\xi}{3\mathcal N} \Big)^{1/4}\, \Big(\frac{\MPl}{H}\Big)^{1/2} \simeq (0.998 - 0.999) \times
 15 \, \Big(\frac{\MPl}{H}\Big)^{1/2}.
 \end{align*}}

If one insists on phenomenologically acceptable ranges of $H \ll M_{\rm Pl}$, e.g. \eqref{Hinfl}, then
\begin{equation}\label{transpl}
 \mu \sim 10^{3} \, M_{s}.
 \end{equation}
 This provides, through \eqref{krbeom2}, a self-consistent and necessary condition for ${\dot b}$ to be approximately constant during inflation,  which thus remains {\it undiluted} at the end of the inflationary period of the string Universe:
 \begin{align}\label{lv}
 \dot{\overline b} = \sqrt{\frac{2}{3}}\, \frac{\alpha^\prime}{96 \, \kappa} \, {\mathcal K}^{0} \simeq {\rm constant}~.
 \end{align}
 {which can be integrated to give:
  \begin{align}\label{lvint}
 {\overline b}(t) = {\overline b}(0) + \sqrt{\frac{2}{3}}\, \frac{\alpha^\prime}{96 \, \kappa} \, {\mathcal K}^{0}\, t~,
 \end{align}
 where $\overline b(0)$ is an initial value of the KR axion field, at the beginning of inflation, immediately after the Big Bang.}  {The value and sign of  $\overline b(0)$  cannot be known at this point, but it will be motivated at due time in our framework when we discuss the vacuum energy density during the inflationary epoch, see Eq.\eqref{lambda}.}

In \cite{bmas,bmasPRD} the assumption that $M_s \sim M_{\rm Pl}$ was made for concreteness, which led to transplanckian graviton modes. We have argued though~\cite{bmasPRD} that this does not present any conceptual problem for our effective theory, but simply implies that the constant Lorentz violating solution for the KR
 background is associated with the primordial graviton waves that are generated deeply inside the quantum gravity region of momenta.

{In view of \eqref{HMs} and \eqref{Hinfl}}, one obtains from \eqref{transpl} the sufficient condition
 \begin {align}\label{murange}
 \mu \gg 2.61 \times (10^{-3} - 10^{-2}) \, M_{\rm Pl},
 \end{align}
 which, in turn, implies that the cutoff scale $\mu$ can be at least of order of $M_{\rm Pl}$, which is a quite natural order of magnitude for the UV completion of the low-energy effective theory. In such a case, \eqref{HMs} implies the following range of the minimal allowed order of magnitude of the string scale
 $M_s \gtrsim 10^{-3} \, M_{\rm Pl}$. Saturating from above $M_s \lesssim \MPl$ we thus obtain the following {approximate range} for the string scale
 \begin{align}\label{msr}
 \MPl \, \gtrsim  \, M_s \gtrsim 10^{-3} \, M_{\rm Pl}~,
 \end{align}
  in order to guarantee the Lorentz-violating solution \eqref{lv} for the KR background.

The reader should note that the {(constant) initial  values} of the anomaly $\mathcal K^0(t=0)$
{and of the KR axion ${\overline b(0)}$}
{\it cannot} be predicted in the context of our effective low-energy theory, because they pertain to the UV-completion of the theory, in our case the full string theory.\footnote{We remark at this point that, independently of our considerations here, it was also pointed out in~\cite{lyth} that the predictions of \cite{stephon} for leptogenesis due to primordial chiral fermions depend heavily on the ultraviolet completion of the theory, given that mainly modes in the deep quantum-gravity/string-theory regime contribute to the lepton asymmetry; moreover, as argued in \cite{lyth,paban}, by performing proper ultraviolet regularization, including higher-than-quadratic-order derivative terms, one may effectively obtain much smaller lepton asymmetry than the one claimed in \cite{stephon}, since the cutoff $\mu$ is effectively replaced by the Hubble constant during the de Sitter phase. Par contrast, in our approach, there are no primordial fermions, and leptogenesis during the radiation era occurs in a completely different way~\cite{bms}, due to the presence of the constant Lorentz Violating axial background of the KR field  \eqref{lv}, whose value can be fixed phenomenologically, to produce sufficient leptogenesis~\cite{bmas,bmasPRD}.} {In principle they might  be determined within microscopic string theory models. For our low-energy field theoretic approach here the prameters ${\overline b(0)}$ and $\mathcal K^0(t=0)$ are
going to be fixed phenomenologically below ({\it cf.} \eqref{slowrollepsi2} and \eqref{b0}, \eqref{b02}.}

\vspace{0.75cm}

{\it Gravitational-Anomaly-induced Inflation through Running Vacuum.}

\vspace{0.5cm}

A slow-roll condition on the KR background is consistent with the Lorentz-Violating background solution \eqref{lv}. On imposing`\cite{bmas,bmasPRD}
\begin{align}\label{slowrollepsi}
 \epsilon \sim \frac{1}{2} \frac{1}{(H M_{\rm Pl})^2}\, {\dot {\overline b}}^2 \sim
 10^{-2},
 \end{align}
 consistent with the {Planck data}~\cite{Planck}, implies
\begin{align}\label{slowrollepsi2}
 {\dot {\overline b}} \sim  \sqrt{2\,\epsilon} \, M_{\rm Pl} \, H \sim  {0.1414} \, M_{\rm Pl} \, H\, {\sim 1.414 \cdot 10^{-5} \, M^2_{\rm Pl}~,}
 \end{align}
{where we used \eqref{Hinfl}. Its integral form \eqref{lvint}} then can be written as
\begin{align}\label{slowrollepsi3}
 {\overline b}(t) \, \sim \, \overline b(0) + \sqrt{2\,\epsilon} \, M_{\rm Pl} \, H\, t\, \sim \overline b(0) + 1.414 \cdot 10^{-5} \, M^2_{\rm Pl}\, t.
 \end{align}
{This determines phenomenologically the anomaly ${\mathcal K}^0(0)$:
\begin{align}\label{k0}
{\mathcal K}^0(0) \sim  0.00166 \, M_s^2 \, M_{\rm Pl}
\end{align}
which, on account of \eqref{msr} lies in the range
\begin{align}\label{k0range}
1.66 \cdot 10^{-3}  \gtrsim \frac{{\mathcal K}^0(0)}{\MPl^3}  \gtrsim  1.66 \cdot 10^{-9}.
\end{align}
We shall come back to the phenomenologically acceptable range of ${\overline b}(0)$ later on ({\it cf.} \eqref{b0}, \eqref{b02} below).}

For now, {we come back to the anomalous conservation law} Eq.~\eqref{csder}. We assume a non-zero vacuum expectation value ({VEV}) \eqref{rrt} of the anomaly term, due to gravitational waves,
and assume an isotropic and homogeneous temporal component of the Cotton tensor $\mathcal C^{00}(t)$. Anticipating the latter to be {proportional to $|\Theta|\ll 1$} ({\it cf.} \eqref{rrt})), one obtains from \eqref{csder}, in a mean field approximation, to lowest order in a perturbative $\Theta$ expansion {(whereby in the left-hand side of the equation we use a (spatially-flat) FLRW background space-time), the following result:}
\begin{align}\label{solution}
C^{\mu 0}_{\,\,\,\,\,\,;\mu} &= \frac{d}{dt} \mathcal C^{00} + 4 H\, \mathcal C^{00}  \simeq - \frac{1}{8} \, \dot{\overline b} \, \langle R^{\alpha\beta\gamma\delta} \, \widetilde R_{\alpha\beta\gamma\delta}\rangle
 \simeq  \, - \frac{1}{8} \sqrt{\frac{2}{3}}\, \frac{\alpha^\prime \, \kappa}{12} \, H \,
\frac{1}{\pi^2} \Big(\frac{H}{M_{\rm Pl}}\Big)^2 \, \mu^4 \, {\dot {\overline b}}^2,
 \end{align}
where we used \eqref{tracecot}, and $\overline b$ denotes the KR background, satisfying \eqref{krbeom2}.
Assuming a (approximately)  constant in time $\mathcal C^{00}$ ({because $H$ itself is approximately constant during inflation}) together with homogeneity and isotropy ({\it i.e}. setting $\mathcal C^{0i} =0$), we find  from \eqref{solution} the consistent solution
\begin{align}
 \mathcal C^{00} \simeq -\epsilon \, \sqrt{\frac{2}{3}}\, \frac{\alpha^\prime \, \kappa}{192} \,
\frac{1}{\pi^2} \, \mu^4 \, \, H^4 \, < \, 0,
\end{align}
where we used \eqref{slowrollepsi2} keeping, though, the slow-roll parameter $\epsilon$ generic for the moment.
From \eqref{einsteincs}, this contributes to the energy density of the vacuum a {\it negative} term,\footnote{For the benefit of the reader, we note that the negativity of $\mathcal C^{00}$ is robust against a change of signature of the coefficient of the g$\mathcal C\mathcal S$ term in \eqref{sea4}, given that the latter will be compensated by a corresponding change of signature of the
 averaged anomaly \eqref{rrt}, which is proportional to that coefficient.} in a similar spirit to the Gauss-Bonnet-dilaton coupling~\cite{kanti} which, like the gravity-anomaly term, also involves terms quadratic in the Riemann curvature  tensor:
\begin{align}\label{envac}
\rho^{\rm g\mathcal C\mathcal S} &= \sqrt{\frac{2}{3}}\, \frac{\alpha^\prime}{12\, \kappa}\, \mathcal C^{00} \simeq  -
\frac{2}{3}  \, \frac{1}{\pi^2 \times 192 \times 12} \,  \epsilon \, \Big(\frac{\mu}{M_s}\Big)^4\, H^4 \nonumber \\ & \simeq -
2.932 \times 10^{-5} \,  \epsilon \, \Big(\frac{\mu}{M_s}\Big)^4\, H^4 \, <\, 0.
\end{align}
Using \eqref{A=0}, we then obtain in order of magnitude\footnote{An important remark we would like to make is that the condition \eqref{A=0}
is assumed to be valid as an order of magnitude estimate, and does {\it not} imply that the cutoff $\mu$ varies with $H$ as $H^{-1/2}~$. The quantity $\mu$ is independent of $H$ and a constant in time. This implies that the g$\mathcal C\mathcal S$ term varies as $H^4$, in contrast to the $\rho^b$ term that varies as $H^2$. However, for our solution under which \eqref{A=0} is valid, both terms are of the same order of magnitude.}

\begin{align}\label{envac2}
\rho^{\rm g\mathcal C\mathcal S}  \simeq  - 1.484 \,  \epsilon \, \MPl^2 \, H^2,
\end{align}

From \eqref{cons}, and the first equality of \eqref{solution}, we also obtain
\begin{align}\label{cons3}
&\frac{d}{dt}(\rho^b + \rho^{g\mathcal C\mathcal S}) + 3 H \Big( (1+w_b)\, \rho^b + \frac{4}{3}\rho^{g\mathcal C\mathcal S} \Big)  \simeq 0
\nonumber \\ & \Rightarrow \quad
\rho^b \simeq -\frac{2}{3}\rho^{g\mathcal C\mathcal S}~,
\end{align}
where the last result holds if $\frac{d}{dt}(\rho^b + \rho^{g\mathcal C\mathcal S}) \simeq 0$ ({which is valid to the extent that the expansion rate remains constant during inflation}) and we took into account that the  {EoS of the pure kinetic $b$-fluid (with no potential) is that of stiff matter}, i.e. $w_b=1$. Thus, we see from \eqref{cons3} that the negative value of the $\rho^{g\mathcal C\mathcal S}$ is essential for the consistency of the approach, since
it is only then that the energy conservation of the total stress energy tensor \eqref{cons} leads to consistent results, given the positivity of $\rho_b$. From \eqref{envac2} and \eqref{cons3} we then obtain
\begin{align}\label{ben}
\rho^b \simeq 0.9895 \, \epsilon \, \MPl^2 \, H^2.
\end{align}
The KR axion stress tensor $T^{\mu\nu}_b$ in \eqref{einsteincs}, on the other hand, will contribute $H^2$ terms to the vacuum energy density~\cite{bmas,bmasPRD} (but of the {\it same order of magnitude} as the $\sim H^4$ terms of the gravitational anomaly, due to \eqref{cons3}):
\begin{align}\label{enpressphib2}
 \rho^{b} = \frac{1}{2} (\dot {\overline b})^2 \simeq  \epsilon \, M_{\rm Pl}^2 \, H^2~,
 \end{align}
where we used the first equality in \eqref{slowrollepsi2}. Comparing with \eqref{ben} we can then see the consistency of our approach,
for every value of the slow roll parameter $\epsilon < 1$ and every value of $H$. We can then adopt the range of values for these parameters dictated by the data~\cite{Planck}, \eqref{slowrollepsi} and \eqref{Hinfl}, respectively. The 1$\%$ discrepancy between \eqref{enpressphib2} and \eqref{ben} is to be expected, according to our previous discussion ({ cf. footnote 5}, which implies that the result for \eqref{envac2} for $\rho^{g\mathcal C\mathcal S}$  should be mulitplied by an uncertainty factor  $(1 - \frac{\xi}{3\mathcal N})$, which lies
in the range $0.9889 \lesssim (1 - \frac{\xi}{3\mathcal N}) \lesssim 0.9905$). This is perfectly justified when taking into account also theoretical uncertainties in our estimate \eqref{rrt} of the gravitational-anomaly condensate.

However, as follows from \eqref{envac}, \eqref{cons3}, the total vacuum energy density turns out to be {\it negative}
\begin{align}\label{negative}
\rho_b + \rho^{g\mathcal C\mathcal S}  = \frac{1}{3}\, \rho^{g\mathcal C\mathcal S} \, \simeq - 0.496  \, \epsilon\, \MPl^2 H^2 < \, 0,
\end{align}
indicating that the anomaly induces an instability in the de Sitter vacuum.

However, this is {\it not} the case. Indeed, in our analysis so far, we have assumed the Einstein equations \eqref{einsteincs}, as they follow from the metric variation of the effective action \eqref{sea4}, and then, we have averaged the modified stress tensor over the (quantum)
gravitational-wave perturbations $\langle T^{b+g\mathcal C\mathcal S}_{\mu\nu}\rangle$. The more correct approach is to average the
partition function over gravitational perturbations about a de Sitter background corresponding to the effective action \eqref{sea4}, and then arrive at the corresponding semiclassical equations with respect to the gravitational field.
Equivalently, we may expand the g$\mathcal C\mathcal S$ term in \eqref{sea4} {around the VEV of the operator $\overline b \, R_{\mu\mu\rho\sigma}\, \widetilde R^{\mu\nu\rho\sigma}$, i.e. about the  condensate  induced by averaging over gravitational-wave perturbations of the metric tensor:}
\begin{align}\label{order}
g\mathcal C\mathcal S & = \sqrt{\frac{2}{3}}\, \frac{\alpha^\prime}{96 \, \kappa} \, \int d^4x \, \sqrt{-g}\,  \Big( \langle \, \overline b(x) \,  R_{\mu\nu\rho\sigma}\, \widetilde R^{\mu\nu\rho\sigma} \, \rangle
+ \,\mathbf{ :}\,   b(x) \, R_{\mu\nu\rho\sigma}\, \widetilde R^{\mu\nu\rho\sigma} \, \mathbf{:} \Big)~,
\end{align}
where $\mathbf{:} \dots \mathbf{:}$ denotes proper quantum ordering of (quantum field) operators, which, in the path-integral language, is interpreted as indicating terms with the appropriate subtraction of the UV divergencies, via regularization by means of the UV cut-off $\mu$. This quantum-ordered term can give rise (via its variation with respect to the gravitational field) to a quantum-ordered Cotton tensor \eqref{cotton}, which is traceless  ({\it cf.} \eqref{tracecot}).

On the other hand, the first (averaged) term on the right-hand side of \eqref{order}, i.e. the condensate induced by the anomaly term, will correspond to
an {\it extra term} in the effective action, of the form of an {induced (positive) cosmological constant, which should be added to \eqref{sea4}.  The  vacuum action {term} associated {with} that condensate {reads}: }
{
\begin{align}\label{lambda}
{\mathcal S_{\Lambda\,cond}}  &= \sqrt{\frac{2}{3}}\,
\frac{\alpha^\prime}{96 \, \kappa} \, \int d^4 x \sqrt{-g}\, \langle \overline b \, R_{\mu\mu\rho\sigma}\, \widetilde R^{\mu\nu\rho\sigma} \rangle=
\int d^4 x \, \sqrt{-g}\, \frac{1}{3\pi^2\times 6\times 96}\,\left(\frac{\mu}{M_s}\right)^4 \, \, \sqrt{2\, \epsilon} \,
\Big[\frac{\overline b(0)}{\MPl} + \sqrt{2\, \epsilon} \,  \mathcal N \Big] \, H^4
\nonumber \\ & \simeq   \int d^4 x \, \sqrt{-g}\, \Big(5.86 \times 10^7 \, \, \sqrt{2\, \epsilon} \,
\Big[\frac{\overline b(0)}{\MPl} + \sqrt{2\, \epsilon} \,  \mathcal N \Big] \, H^4 \Big) \,
\equiv  {-} \int d^4x \, \sqrt{-g} \, {\rho_{\Lambda\,cond}}~,
\end{align}}
{where the negative sign in front of the integral on the right-hand side of the above equation is due to our conventions, in which a de Sitter vacuum energy corresponds to a (constant) positive $\rho_{\Lambda\,cond}$. Thus we must have $\overline b(0) <0$ so as to get positive vacuum energy density capable of triggering inflation.  The symbol $\simeq $ indicates an order of magnitude estimate, and we used \eqref{k01},  \eqref{transpl}, \eqref{lv} and \eqref{slowrollepsi3}.} We took also into account that $(H\, \overline t)_{\rm max}$ is a maximum order of magnitude~\cite{bmas,bmasPRD} evaluated at the end of the inflationary period, for which $H\, t_{\rm end} \sim {\mathcal N} = 60-70 = (H\, \overline t)_{\rm max}$, with ${\mathcal N}$ the number of e-foldings. In the above we take $\epsilon \sim 10^{-2}$, as required by inflationary phenomenology ({\it cf.} \eqref{slowrollepsi}). We next notice that, if we consider transplanckian values for $|\overline b(0)| \gg \MPl $ (in analogy with what happens with the inflaton field in conventional inflationary scenarios),
with  {$\overline b(0) <0$},
then the quantity ${\rho_{\Lambda\,cond}> 0}$ in \eqref{lambda} does not change order of magnitude during the entire inflationary period, for which $H \simeq $ constant, and thus it can be approximated by a constant, {leading to a de Sitter situation}. In fact, for this purpose,  {it suffices to assume
\begin{align}\label{b0}
|\overline b(0)| \gtrsim \sqrt{2\, \epsilon}\, {\mathcal N} \, \MPl~,
\end{align}
say,
\begin{align}\label{b02}
|\overline b(0)|\sim 10 \, \MPl.
\end{align}}

{The alert reader might worry about the compatibility of the transplanckian values of the KR background axion, \eqref{b0}, \eqref{b02}, with the validity of the effective low-energy field theory
below Planck scale. However, during the inflationary phase,}  {with the exception of the condensate term \eqref{lambda}}
{our effective theory depends only on derivatives of the massless KR axion, $\dot b$, which in view of \eqref{slowrollepsi2}, is sufficiently smaller than $\MPl^2$ to justify the validity of the effective field theory \eqref{sea4}, and also ignoring higher order in $\alpha^\prime$ corrections, as discussed in the Appendix.}
{Moreover, as we show below ({\it cf.} \eqref{toten}), the condensate term \eqref{lambda} itself, being proportional to $H^4$, also assumes sub-Planckian values, despite the Planck size fluctuations of the KR field, compatible with the validity of the effective theory and with current phenomenology.}

The reader should take notice of the fact that, as typical with other condensates in field theory, {\it e.g.} gluon condensates in QCD, the {gravitational-anomaly condensate  $\langle \overline b (x) \, R_{\mu\mu\rho\sigma}\, \widetilde R^{\mu\nu\rho\sigma} \rangle $
 in \eqref{lambda}
is an (scalar) invariant which does not depend on the metric tensor, whereby it may lead to a
positive-cosmological-constant (de Sitter) type term in the effective action under the above mentioned conditions}.  In a sense, the term \eqref{lambda} is equivalent to a quantum-gravity-induced ``trace'' of the Cotton tensor, which, as we have seen above, is {\it classically} traceless \eqref{tracecot}. Such a $\Lambda$-type-term cannot arise in a classical general-relativistic treatment, and, hence, it was not considered in the analysis of \cite{jackiw}. {Notice also that the quantum induced ${\rho_{\Lambda\, cond}}$ term is approximately constant during the de Sitter stage, but it evolves with time (dynamical vacuum energy) in subsequent eras, as is characteristic of the RVM-type models\,\cite{RVMreviews}.}

The cosmological constant type term \eqref{lambda}, then, leads to an additional $\Lambda$-de-Sitter-type induced contribution to the modified stress-energy tensor \eqref{cons}, {with the characteristic EoS of vacuum $\rho_{\Lambda\,cond} = -p_{\Lambda\,cond} $.  Such EoS is maintained even when the vacuum energy density becomes dynamical, as the energy-momentum tensor still keeps the pure vacuum form proportional to the metric.
The presence of $\rho_{\Lambda\,cond}$  still preserves the conservation law  \eqref{cons}, and corresponds to the following (positive) contribution to the total energy density:
${\rho_{\Lambda\,cond}} \sim   10^8 \, \sqrt{2\, \epsilon} \, \frac{|\overline b(0)|}{\MPl} \, H^4$,}
which, for $\epsilon \sim 10^{-2}$, $\mathcal N ={\mathcal O}(60-70)$ and ${|\overline b(0)|}\gtrsim 10 \, \MPl$ {({\it cf.} \eqref{b0})}, dominates the total energy density,
\begin{align}\label{toten}
\rho_{\rm total} = \rho_b + \rho_{g\mathcal C\mathcal S} + {\rho_{\Lambda \, cond}} \simeq
3\MPl^4 \, \Big[{ -1.65} \times 10^{-3} \Big(\frac{H}{\MPl}\Big)^2
+ \frac{\sqrt{2}}{3} \, \frac{{|\overline b(0)|}}{\MPl}\, \times {5.86\, \times} \, 10^6 \, \left(\frac{H}{\MPl}\right)^4 \Big] > 0~.
\end{align}
{This expression, {by virtue of equations (\ref{Hinfl}) and \eqref{b0}}),  is {\it positive} and drives the de Sitter (inflationary) space-time.}

Let us now compare the expression  \eqref{toten} with the form of the energy density of the so-called ``running vacuum model'' (RVM) of the Universe~\cite{ShapSol}, according to which the vacuum energy density of the Universe, after integrating out
matter degrees of freedom, reads:
\begin{equation}\label{rLRVM}
\rho^{\Lambda}_{\rm RVM}(H)
= \frac{3}{\kappa^2}\left(c_0 + \nu H^{2} + \alpha
\frac{H^{4}}{H_{I}^{2}}\right) + \dots \;,
\end{equation}
where the coefficients $\nu$ and $\alpha$ are constants, $H_I$ is the Hubble parameter close to GUT scale, and $c_0$ is an integration constant, which in the early Universe is not dominant, while it ({approximately}) coincides with the cosmological constant in the late Universe ({up to a correction of ${\cal O}(\nu)$}). For the conventional RVM, the expectation is that {both  $\nu$ and $\alpha$ are positive}~\cite{ShapSol}. The $\dots$ denote terms of higher order in $H^2$ (due to general covariance the expansion is necessarily in terms of even powers of the Hubble parameter $H$).
As mentioned in the introduction, the model is in agreement with observations~\cite{rvmpheno}. The $\sim H^4$  terms in \eqref{rLRVM} are {\it not suppressed} by heavy masses, and although irrelevant for the current universe, nonetheless they can play a central r\^ole in the early universe and can explain inflation and successful exit from it ({see ~\cite{rvmpheno,bamasol,Yu2019} for details}), without the need for introducing an external inflaton field. {The $H^4$ terms characteristic of  the RVM} are equivalent to the presence of a slowly-rolling internal scalar degree of freedom,
which in the scenario discussed in \cite{bamasol} and here, is provided by the scalar mode hidden in the quantum fluctuations of the graviton condensate.

On comparing \eqref{toten} with \eqref{rLRVM}, by identifying $\rho_{\rm total}$ and $\rho^{\Lambda}_{\rm RVM}(H)$, we make the following observations for our model:
\begin{itemize}
\item{(i)} In our string-inspired model for the early Universe we have $c_0=0$.  Such a term may appear in the late eras of the Universe, e.g. through the generation of a potential for the $b(x)$ field~\cite{bmas,bmasPRD}.

\item{(ii)}
As a result of the negative contributions of the Cotton tensor
to the energy density $\rho_{\rm total}$, the coefficient of the $H^2$ terms in \eqref{toten} would imply, on account of \eqref{rLRVM}, a $\nu < 0$ in the early Universe, where gravitational anomaly contributions dominate. However there is no contradiction with the spirit of the RVM. Indeed, in  our case, the Cotton tensor is {\it not} a vacuum contribution, as it is associated with {\it gravitational-wave excitations} on the FLRW metric background space-time. For the background space-time the Cotton tensor vanishes, as we have already mentioned~\cite{jackiw}.
On the other hand, the KR axion is associated with the spin-one antisymmetric tensor field of the massless gravitational multiplet of strings~\cite{string}, which in the case of the (phenomenologically relevant) superstring constitutes the ground state, due to the absence of tachyon modes from the spectrum.
In this sense, the RVM should be associated with the contributions of the $b$-axion field stress tensor $T^{\mu\nu}_b$ \eqref{tbstress} alone, ignoring the Chern-Simons terms, which, on account of \eqref{slowrollepsi}, \eqref{slowrollepsi2} leads ({\it cf.} \eqref{rLRVM}) to a {\it positive} $\nu$ coefficient ({which we denote $\nu_b$}), given in our framework by
{
\begin{equation}\label{nub} \nu_b\equiv\frac{\kappa^2\,\dot{b}^2}{6H^2}= \frac{\epsilon}{3}\simeq 3 \times 10^{-3} > 0\,,
\end{equation}}
\noindent which emerges from  the corresponding expression for $\rho_b = T_{b\,0}^{0}$~\cite{bmas,bmasPRD}. {Let us notice that this theoretical prediction is just within the right order of magnitude of the typical fitting values obtained for the $\nu$-parameter using the wealth of observational data on CMB, BAO, LSS and expansion history, see the detailed analyses\,\cite{rvmpheno}. This is realistic and remarkable, as this is the situation expected in the present era in which the effect of the  Chern-Simons terms are irrelevant, only to reappear in the remote future when the final de Sitter phase will become fully dominant anew. }

{We stress that the positivity of $\nu_b$ in the radiation and matter dominated eras is warranted, since the gravitational anomalies cancel during the standard FLRW periods~\cite{bmas,bmasPRD}. In particular, in the late epochs we find contributions to the Universe energy density of order $H^2$ due to the late-era KR axion field (whose background configuration receives contributions from chiral anomalies~\cite{bmas,bmasPRD}). These come with a {\it positive} $\nu > 0$ (which has been argued~\cite{bmas,bmasPRD} to be of the same order as the primordial $\nu_b$ above \eqref{nub}, on phenomenological reasons). This result, which as indicated is phenomenologically sound\cite{rvmpheno}, is also in remarkable agreement with the theoretical expectations based on general renormalisation-group arguments for the RVM\,\cite{Fossil07,RVMreviews}.}

\item{(iii)} On the other hand,
we find  that the coefficient $\alpha $ is {\it positive} already during the inflationary era, and of order:
\begin{equation}\label{eq:alphavalue}
\alpha= \frac{\sqrt{2}}{3} \, \frac{{|\overline b(0)|}}{\MPl} \, \times \, {5.86\, \times} \,10^6 \left (\frac{H_I}{\MPl}\right)^2\sim {2.8 \times 10^{-2}} \, \frac{{|\overline b(0)|}}{\MPl}~,
\end{equation}
assuming a (typical) Hubble parameter $H_I$ during inflation of order \eqref{Hinfl}.
Notice that the value of $\alpha$ does not depend on the specific magnitude of the string scale, but only on the ratio
$\mu/M_s$, {as follows from \eqref{A=0}, used in the estimate of the total energy density \eqref{toten}}. From \eqref{toten}, and  \eqref{Hinfl}, then, one easily sees that we may identify the total energy density with a GUT-like potential $V \sim M_X^4$ corresponding to an energy scale $M_X$:
\begin{align}\label{mxscale}
\rho_{\rm total} &\simeq {\rho_{\Lambda\, cond}}  \sim M_X^4 \simeq  \sqrt{2} \, \frac{{|\overline b(0)|}}{\MPl}\, {5.86}\times10^{-10}\, \MPl^4\ {\simeq \frac{{|\overline b(0)|}}{\MPl}\, 8.3\times10^{-10}\, \MPl^4}
\nonumber \\&  \Rightarrow \quad M_X {\simeq 1.3 \, \times 10^{16}\,  \Big(\frac{{|\overline b(0)|}}{\MPl} \Big)^{1/4}~ {\rm GeV} \simeq 2.3 \, \times 10^{16}~ {\rm GeV}\,,}
\end{align}
{for ${|\overline b(0)|} \gtrsim 10 \, \MPl$, as indicated before. As it turns out, the GUT scale that we associate to the total energy density in the early epoch  is in perfect agreement with generic RVM predictions based on GUT models~\cite{RVMreviews}.}

\end{itemize}

\vspace{0.5cm}

{\it Insensitivity of the results to specifics of UV Completion.}

\vspace{0.5cm}

At this point, we would like to offer support to the {insensitivity of our findings to the specifics of UV completion}, by demonstrating that the Lorentz-Violating constant KR backgrounds \eqref{krbeom2}, \eqref{lv} constitute solutions to the axion equations of motion obtained from the generic one-loop effective action of \cite{paban}, not necessarily in the context of string theory.
Indeed, for a generic g$\mathcal C\mathcal S$ coupling (in the notation of \cite{paban})
$\int d^4 x \sqrt{-g}\, {\cal F } (\tilde a) R_{\mu\nu\rho\sigma} \, \widetilde R^{\mu\nu\rho\sigma} $, where $\tilde a$ denotes a (generic) pseudoscalar field,  the one-loop effective action obtained by integrating out graviton fluctuations about a given background space time $\hat g_{\mu\nu}$ is given by
\begin{align}
 W[{\cal F}] &={\MPl}^2  \int \,\, d^4 x  \sqrt{-\hat{g}}   \Big[ \hat{R} +   a_1 \, ( \partial_{\mu} {\cal F} \, {\hat g} ^{\mu \nu}  \partial_{\nu} {\cal F} ) \,
   +    a_2\,  \frac{1}{{\MPl}^2} \,{ \hat R}^2 + a_3 \,\frac{1}{{\MPl}^2} \, {\hat R}_{\mu \nu} {\hat R}^{\mu \nu}  \nonumber \\
 &+  ( a_4 \, \frac{{ \hat R }}{{\MPl}^2} \,  {\hat g} ^{\mu \nu} +  a_5 \, \frac{ {\hat R} ^{\mu \nu}}{{\MPl}^2}) \, ( \partial_{\mu} {\cal F} \, \partial_{\nu} {\cal F} )   +    a_6 \,\frac{1}{{\MPl}^2} \, ( \partial_{\mu} {\cal F} \,   {\hat g} ^{\mu \nu}   \partial_{\nu} {\cal F} ) ^2  \, +  a_7 \, \frac{1}{{\MPl}^2} \,( \hat \Box {\cal F} ) ^2      +      \dots       \Big] \label{effa}
\end{align}
where the $\dots$ denote higher derivative terms, and hatted quantities denote the ones pertaining to a metric background, in our case taken to be the inflationary (de Sitter) FLRW space-time. The symbol $\hat \Box$ denotes the covariant d'Alembertian in the background (de Sitter) space-time. The (dimensionless) coefficients $a_i$, $i=1, \dots 7$ depend on the specific UV completion, and in general are divergent, thus becoming  functions of the UV cutoff $\mu$ after proper regularisation (such regularised coefficient are background independent~\cite{paban}). Not all terms of \eqref{effa} are independent, as they can be related by field redefinitions, but this is not of specific interests to us.
In our specific string-inspired case ({\it cf.} \eqref{sea4}) ${\mathcal F} =  \sqrt{\frac{2}{3}}\,
\frac{\alpha^\prime}{96 \, \kappa} \, b(x) $.

Assuming a homogeneous and isotropic ${\mathcal F}(t)$ field,  and an inflationary space-time de Sitter background, for which
$\hat R_{\mu\nu} =3 H^2 \hat g_{\mu\nu}$ with $H = $ constant, we shall seek
solutions to the equations of motion for the pseudoscalar field $\mathcal F(t)$ which are of the Lorentz-violating form \eqref{krbeom2},\eqref{lv}:
\begin{align}\label{fconst}
{\dot{\cal F}} = {\rm  constant}~.
\end{align}
Under those circumstances, the pertinent equations read:
\begin{align}\label{uvsol}
&\partial_t \Big(\sqrt{-\hat g} \Big[ 2a_1 \MPl^2  + 2 \, H^2 \,(12 a_4 + 3a_5)   + 2 \, a_6 \, \frac{1}{\MPl^2} \Big({\dot{\cal F}}\Big)^2 \,
+ \dots \Big]  \, {\dot{\cal F}} \Big)=0,
\end{align}
where the $\dots$ denote the corresponding terms obtained from the last term on the right-hand side of \eqref{effa}, with coefficient $a_7$, which for solutions of the type \eqref{fconst}, we are seeking for, vanish. From \eqref{uvsol}, we indeed observe that there can be solutions of the type \eqref{fconst}, if
\begin{align}\label{uvsol2}
2a_1 \MPl^2 + 2 \, H^2 \,(12 a_4 + 3a_5)  + 2 \, a_6 \, \frac{1}{\MPl^2} \Big({\dot{\cal F}}\Big)^2 =0,
\end{align}
which implies a solution of the type \eqref{fconst}, provided some of the $a_i$ coefficients in \eqref{effa} are negative, something which is generically expected, due to the fact that UV subtractions have taken place in order to obtain \eqref{effa}.

This completes our argument on the robustness of the existence of Lorentz-violating solutions \eqref{fconst}, or \eqref{krbeom2},\eqref{lv}, {independently of the specifics of the} UV completion of effective theories in a generic framework.  {We remind the reader that taking into account terms of higher order in $\alpha^\prime$ that appear in a string effective action} does not affect our conclusion on the existence of the solution \eqref{lv}, as argued in the Appendix.}
Of course, to obtain the value of the solution one needs to have a full knowledge of the underlying UV complete
quantum gravity model, e.g. the full string theory in our case, something which may not be available.

\vspace{0.5cm}

{\it Post Inflationary Era Chiral Anomalies and  Axion Dark Matter.}

\vspace{0.5cm}

In the cosmological model of \cite{bmas,bmasPRD}, it was assumed that at the end of inflation
the appearance of chiral fermions, with anomalous axial currents, among other matter occurs.  Then, in models involving right-handed massive neutrinos,  matter-antimatter asymmetry in the observable universe could be due to such an anomaly in the post-inflationary era through the mechanism advocated in~\cite{bms}, as a consequence of the appearance of the undiluted KR background \eqref{krbeom2},\eqref{lv}. For details we refer the reader to \cite{bmasPRD}.

The effective action of chiral fermions during the post inflationary eras is crucially based on
 the link of the KR axion $b(x)$ with the torsion provided here~\cite{string,kaloper} by the (totally antisymmetric) quantity $\epsilon_{\mu\nu\rho\sigma} \partial^\sigma b$, which is dual to the Kalb-Ramond antisymmetric tensor field strength ${\mathcal H}_{\mu\nu\rho}$, as discussed previously ({\it cf}. \eqref{torcon}). Indeed, such a torsion is present in the gravitational covariant derivative of the fermion Dirac term, which leads eventually to
the coupling of the axial fermion current with the KR axion field $b(x)$. The  effective action was derived in~\cite{bmas,bmasPRD}, and reads:
\begin{align}\label{sea6}
S^{\rm eff} &=\; \int d^{4}x\sqrt{-g}\Big[ -\dfrac{1}{2\kappa^{2}}\, R + \frac{1}{2}\, \partial_\mu b \, \partial^\mu b   -  \sqrt{\frac{2}{3}}\,
\frac{\alpha^\prime}{96\, \kappa} \, \partial_\mu b(x) \, {\mathcal K}^\mu
\Big]  \nonumber \\
& + S_{Dirac}^{Free} + \int d^{4}x\sqrt{-g}\,  \frac{\alpha^\prime}{\kappa} \, \sqrt{\frac{3}{8}} \, \partial_{\mu}b \, J^{5\mu}    - \dfrac{3\, {\alpha^\prime}^{2}}{16\, \kappa^2}\, \int d^{4}x\sqrt{-g}\,J^{5}_{\mu}J^{5\mu}  + \dots,
\end{align}
where $J_\mu^5$ denotes the (anomalous in general) fermion axial current, summed over all fermion species in the model, and the $\dots$ indicate gauge field kinetic terms, as well as terms of higher order in derivatives, of no direct relevance to us here. The reader should notice the four fermion axial-current-current term in \eqref{sea6}, which is characteristic of Einstein-Cartan theories with torsion~\cite{hehl,shapiro}.

In the scenario of \cite{bmas,bmasPRD} the generation of {\it chiral matter} at the end of inflation leads to a {\it cancellation} of the gravitational anomalies {\it locally}, thus restoring  diffeomorphism invariance in the presence of matter, required for consistency of the
matter/radiation quantum field theory. However, {\it chiral}~\cite{adler} or {\it QCD-axion}~\cite{qcdaxion} type {\it anomalies} may  remain {\it uncompensated}. These do not contribute to stress tensor of matter, unlike the gravitational ones, hence there is no fundamental reason for the matter theory to be chiral-anomaly free, only the gauge symmetry must be anomaly free so as to preserve the Ward identities. Thus, we postulate the following relation during the radiation (and matter) eras:
\begin{align}\label{anom2}
& \partial_\mu \Big[\sqrt{-g}\, \Big(  \sqrt{\frac{3}{8}} \frac{\alpha^\prime}{\kappa}\, J^{5\mu}  -  \frac{\alpha^\prime}{\kappa}\, \sqrt{\frac{2}{3}}\,
\frac{1}{96} \, {\mathcal K}^\mu  \Big) \Big]   =   \sqrt{\frac{3}{8}} \, \frac{\alpha^\prime}{\kappa}\, \Big(\frac{\alpha_{\rm EM}}{2\pi}  \, \sqrt{-g}\,  {F}^{\mu\nu}\,  \widetilde{F}_{\mu\nu} + \frac{\alpha_s}{8\pi}\, \sqrt{-g} \, G_{\mu\nu}^a \, \widetilde G^{a\mu\nu} \Big)~,
\end{align}
where $F_{\mu\nu}$ is the electromagnetic (EM) Maxwell tensor,  and $G_{\mu\nu}^a$ is the gluon field strength, with $a=1, \dots 8$ an adjoint SU(3) colour index, $\alpha_{\rm EM}$ is the electromagnetic fine structure constant, and $\alpha_s$ is the strong interactions fine structure constant. The fact that the anomaly is proportional to these {fine}  structure constants is due to the fact that it is a one-loop effect, with chiral fermions circulating in the loop.

In \cite{bmas,bmasPRD} we considered only the effects of the electromagnetic chiral anomalies only, arguing in favour of their r\^ole in the generation of large-scale cosmic magnetic fields, which lead to $H^2$ contributions to the vacuum energy density, which again assumes an RVM form \eqref{rLRVM}. Here we concentrate on the QCD anomalies, which are assumed dominant during the QCD-epoch of the Universe.
By partially integrating the $b-J^5$ interaction term in \eqref{sea6}, and using \eqref{anom2}, it is straightforward to observe that in the QCD era one obtains an effective action for the KR pseudoscalar $b(x)$ of the form
\begin{align}\label{qcd}
S^{\rm eff}_b =&\; \int d^{4}x\sqrt{-g}\Big[  \frac{1}{2}\, \partial_\mu b \, \partial^\mu b -  \frac{\alpha^\prime}{\kappa}\, \, \sqrt{\frac{3}{8}} \, \frac{\alpha_s}{8\pi}\, b(x) \, G_{\mu\nu}^a \, \widetilde G^{a\mu\nu}\Big]~.
\end{align}
We now remark that shift-symmetry-breaking QCD instanton (non-perturbative) effects can generate a periodic potential for the KR axion during the QCD era
\begin{align}\label{vqcd}
V_b^{\rm QCD} & \simeq \Lambda^4_{\rm QCD} \Big(1 - {\rm cos}(\frac{b}{f_b})\Big)~,  \nonumber \\
 f_b &\equiv \sqrt{\frac{8}{3}} \, \frac{\kappa}{\alpha^\prime} =  \sqrt{\frac{8}{3}} \, \Big(\frac{M_s}{\MPl}\Big)^2\, \MPl ~,
\end{align}
minimisation of which fixes the strong-CP-violating angle $\langle \theta_{\rm CP} \rangle =0$. In \eqref{vqcd}, $\Lambda_{\rm QCD} \sim 218~{\rm MeV}$ is the QCD scale, and $f_b$ plays the r\^ole of the (mass-dimension-one) QCD axion decay coupling constant $f_a$, which is estimated phenomenologically to lie in the range~\cite{qcdaxion}
\begin{align}\label{far}
 10^9~{\rm GeV} < f_a < 10^{12}~{\rm GeV}~,
\end{align}
although the upper bound can be extended up to $10^{17}~{\rm GeV}$ by means of astrophysical constraints~\cite{astro}.
In our string-inspired case, this is determined by the string scale $\alpha^\prime = M_s^{-2}$. For the validity of our Lorentz-violating constant solution  for the KR axion \eqref{lv}
we have seen that \eqref{msr} must be valid, which implies a range
\begin{align}\label{fbr}
{3.9 \times 10^{12} ~{\rm GeV}\lesssim\, f_b \,\lesssim\,3.9 \times 10^{18} ~{\rm GeV}\,,}
\end{align}
where we used that $M_{\rm Pl}= 2.4 \times 10^{18}$~GeV. We thus observe that there is a marginal overlap (in order of magnitude) between the minimally allowed region of $f_b$ \eqref{fbr} and the maximally allowed phenomenological region of the QCD axion coupling constant \eqref{far}. If the constraints of \cite{astro}, however, are taken into account, we see that the overlap of the allowed regions between $f_a$ and $f_b$ increases significantly.

The instanton-induced KR-axion mass is then given by
\begin{align}\label{axionmass}
m_b &= \sqrt{\left. \frac{\partial^2 V_b^{\rm QCD}}{\partial b^2}\right|_{b=0}}  = \frac{\Lambda^2_{\rm QCD}}{f_b} = \sqrt{\frac{3}{8}} \, \Big(\frac{\Lambda_{\rm QCD}}{M_s}\Big)^2\,  \MPl
= \sqrt{\frac{3}{8}} \, \Big(\frac{\Lambda_{\rm QCD}}{\MPl}\Big)^2 \, \Big(\frac{\MPl}{M_s}\Big)^2\,  \MPl ~,
\end{align}
which, in view of \eqref{msr}, lies in the range
\begin{align}
{ 1.17 \times 10^{-11} ~{\rm  eV} \lesssim m_b \, \lesssim \, 1.17 \times 10^{-5}~{\rm  eV} \,,}
\end{align}
which lies well within the range calculated in lattice QCD approaches~\cite{latticeqcd}: $m_a \sim 5.7 \, (\frac{10^{12}~{\rm GeV}}{f_a}) \, \times 10^{-6}$~eV.

\vspace{0.5cm}

{The above considerations are rather string-theory-model independent, in the spirit of \cite{dine} and \cite{svwit}, where the KR axion is viewed as a Lagrange multiplier of the modified Bianchi identity \eqref{modbianchi2}, and acquires dynamics by dualization (path-integral integration) of the KR field strength  $\mathcal H$.
Our point above was to present the simplest of the scenarios, in which, during the QCD epoch,
non-perturbative QCD instanton effects in our effective field theory \eqref{qcd} generate a KR axion potential,
and examine whether the non-diluted solution \eqref{lv}, \eqref{slowrollepsi2} for the $b$-axion, resulting from {\it primordial-gravitational-wave condensates}, provides phenomenological consistency for the pertinent axion parameters.
We found a rather marginal agreement with cosmology, see \eqref{fbr}. In terms of microscopic string theory models, discussed in \cite{dine,svwit}, where we refer the interested reader for more information, there is a plethora of different ranges for the axion parameters. In most models, like in our case here, the axion coupling constant $f_a$ is found larger than the GUT mass scale,  {outside its cosmological bounds}, although there are models in which $f_a$ is much smaller. It would be interesting to discuss specific string theory model realisations of the solutions \eqref{lv}, \eqref{slowrollepsi2}, in the spirit of \cite{dine,svwit}. This falls outside our scope. We hope though that our current work and that in \cite{bmasPRD} serve as motivations for such a study in the future.}

\vspace{0.5cm}

{{\it Summary}}

\vspace{0.5cm}

{In this Letter, we have shown that the {leading order structures of the energy density of the running vacuum model (RVM), namely the $H^4$ and $H^2$ terms (the latter being subdominant in the early Universe but playing a role at late epochs)
can be derived from the effective action of string theory}.  The high power $H^4$  is able to produce inflation with graceful exit.  The lower power  $H^2$ carries a negative coefficient ($\nu<0$)  in the early universe, indicating that the anomaly triggers instabilities in the de Sitter vacuum and as a result inflation quickly transmutes into a standard radiation regime. However, {the coefficient $\nu$  flips sign at this point  to a positive one}, since the {gravitational} anomaly contribution disappears during the standard epochs of the FLRW evolution}.

{At  low energy, the sign $\nu>0$ is crucial since it makes the RVM to mimic quintessence-like behavior at present. The theoretically predicted value of $\nu$ for the post-inflationary universe (viz. $\nu=+{\cal O}(10^{-3}$) is nicely in agreement with the existing fitting analyses of the RVM in the light of the modern observational data, as shown in detail in\,\cite{rvmpheno}. At the same time, we have found that in the above simplified scenario,} the KR axion itself becomes the Dark-Matter axion through the generation of a non-perturbative potential in the QCD era, in which case  the spontaneous Lorentz symmetry breaking solution ${\dot b}$ = constant ({\it cf.} \eqref{lv}, \eqref{slowrollepsi2}) ceases to exist. However, as discussed in \cite{bmasPRD}, there are other, more complicated, but in the same spirit, mechanisms for generating mass for the KR axion through non-perturbative potentials arising in string theory models~\cite{axioninfl}
involving {\it mixing} of the $b(x)$ fields with other axions that are abundant in string theory~\cite{arv}. These scenarios are capable of preserving the structure \eqref{lv}, \eqref{slowrollepsi2} for the KR axion background at modern eras, notably with an $\epsilon $ of the same order as the primordial one. They also allow for ultralight axion dark matter with mass less than $10^{-21}$~eV, which constitutes currently the subject of intense research~\cite{sensors}.  {Overall, the theoretical framework presented here suggests that string-inspired RVM models can provide  a {global explanation for inflation,  dark matter (of axionic nature) and (dynamical) dark energy} in the form of running vacum.}

\section*{Acknowledgements}

{We thank the referee for instructive comments and remarks on string theory considerations, which
helped improving the presentation.}
SB acknowledges support from
the Research Center for Astronomy of the Academy of Athens in the
context of the program  ``{\it Tracing the Cosmic Acceleration}''.
The work  of NEM is supported in part by the UK Science and Technology Facilities  research Council (STFC) under the research grants
ST/P000258/1 and ST/T000759/1. The work of JS has
been partially supported by projects  FPA2016-76005-C2-1-P (MINECO), 2017-SGR-929 (Generalitat de Catalunya) and MDM-2014-0369 (ICCUB).
This work is also partially supported by the COST Association Action CA18108 ``{\it Quantum Gravity Phenomenology in the Multimessenger Approach (QG-MM)}''.
NEM acknowledges a scientific associateship (``\emph{Doctor Vinculado}'') at IFIC-CSIC-Valencia University, Valencia, Spain.

\section*{{Appendix: String Theory Considerations}}

In this Appendix we would like to justify the use of the effective action \eqref{sea2}, based only on antisymmetric tensor and graviton degrees of freedom. What we shall argue below is that a constant (or slowly moving) dilaton
configuration, as assumed above and in \cite{bmasPRD},  can be consistently implemented, and all our conclusions
are not affected by the inclusion of the dilaton dynamics, upon certain reasonable assumptions that we shall outline explicitly.

Our starting point is the string effective action of the massless string gravitational multiplet (graviton, dilaton and
KR antisymmetric tensor fields)
in the Einstein frame~\cite{gsw,string,kaloper}~\footnote{Our conventions and definitions used throughout this work are: signature of metric $(+, -,-,- )$, Riemann Curvature tensor
$R^\lambda_{\,\,\,\,\mu \nu \sigma} = \partial_\nu \, \Gamma^\lambda_{\,\,\mu\sigma} + \Gamma^\rho_{\,\, \mu\sigma} \, \Gamma^\lambda_{\,\, \rho\nu} - (\nu \leftrightarrow \sigma)$, Ricci tensor $R_{\mu\nu} = R^\lambda_{\,\,\,\,\mu \lambda \nu}$, and Ricci scalar $R = R_{\mu\nu}g^{\mu\nu}$.}
\be\label{sea}
S_B  =\; \int d^{4}x\sqrt{-g}\Big( \dfrac{1}{2\kappa^{2}} [-R + 2\, \partial_{\mu}\Phi\, \partial^{\mu}\Phi] - \frac{1}{6}\, e^{-4\Phi}\, {\mathcal H}_{\lambda\mu\nu}{\mathcal H}^{\lambda\mu\nu} - V(\Phi) \Big) + S_B^{(\alpha^\prime)} ,
\ee
where  $V(\Phi)$  is a (non-perturbartively) generated potential for the the dilaton field $\Phi$, which we leave unspecified for the purposes of this work.

The $S_B^{(\alpha^\prime)} $ represent two classes of string low-energy effective action terms, of higher order in $\alpha^\prime$~\cite{string}, with $\alpha^\prime = M_s^{-2}$ the Regge slope of the string and $M_s$ the string mass scale, which is not necessarily the same as the four dimensional gravitational constant $\kappa^2 = 8\pi \, {\rm G} = M_{\rm Pl}^{-2}$: (i) higher (than two) derivative terms,
including couplings of dilatons to the Gauss-Bonnet quadratic curvature invariant
and (ii) terms involving higher than quadratic powers of ${\mathcal H}_{\mu\nu\rho}$. Their explicit form to $\mathcal  O(\alpha^\prime)$ reads~\cite{string}:
\begin{align}\label{seaPhi}
S_B^{(\alpha^\prime)}  &= \;  \frac{ \alpha^\prime}{g_s^{(0)2}\, \kappa^2} \, \int d^{4}x\sqrt{-g}\Big[- c_1  \frac{1}{8} e^{-2\Phi}
 \Big(R_{\mu\nu\rho\sigma} \, R^{\mu\nu\rho\sigma} - 4 R_{\mu\nu}\, R^{\mu\nu} + R^2\Big)
 + c_2 e^{-2\Phi} (\partial_\rho \Phi \,\partial^\rho \Phi)^2 + \nonumber \\
  &+ c_3 \, e^{-2\Phi}\,(R^{\mu\nu} -\frac{1}{2} \, g^{\mu\nu} \, R)\,  \partial_\mu \Phi \, \partial_\nu \Phi
 + c_4 \, \, e^{-2\Phi}\,\nabla^\mu \partial^\nu \Phi \, \partial_\mu \Phi \, \partial_\nu \Phi \nonumber \\
 &+
 \;  c_5\, e^{-6\Phi}\,(-\nabla^\mu \mathcal H^{\nu\rho\sigma} \nabla_\nu \mathcal H_{\sigma\mu\rho} +
 \nabla_\mu \mathcal H^{\mu\rho\sigma} \, \nabla^\nu \mathcal H_{\nu\rho\sigma} ) \nonumber \\
& + c_6 \, e^{-6\Phi}\,( 5R^{\mu\nu\rho\sigma}\,  \mathcal H_{\mu\nu\lambda}\, \mathcal H^\lambda_{\,\,\rho\sigma}
 -8R^{\mu\nu}\, \mathcal H_{\mu\rho\sigma} \,\mathcal H_\nu^{\,\,\rho\sigma} + R\,\mathcal H_{\mu\nu\rho}\, \mathcal H^{\mu\nu\rho} ) \nonumber \\
&+ c_7 \, e^{-6\Phi}\,\mathcal H^{\mu\nu\rho} \, \partial_\rho \Phi \, \partial^\lambda \Phi \, \mathcal H_{\lambda\mu\nu}
+c_8 \, e^{-6\Phi}\,\mathcal H_{\mu\nu\rho} \, \mathcal H^{\mu\nu\rho} \, \partial_\lambda \Phi \, \partial^\lambda \Phi
\nonumber \\
& + c_9 \, e^{-10\Phi}\,\Big(-3\mathcal H^{\mu\nu\rho} \, \mathcal H_{\lambda\nu\rho} \, \mathcal H^{\lambda\sigma\tau} \mathcal H_{\mu\sigma\tau} +  \mathcal H^{\mu\nu\rho} \, \mathcal H_{\mu}^{\,\,\lambda\sigma}  \, \mathcal H_{\nu\lambda}^{\,\,\,\,\,\,\kappa}\mathcal H_{\kappa\rho\sigma}  + \frac{2}{3} (\mathcal H_{\mu\alpha\beta}\, \mathcal H_{\nu\alpha\beta} )^2 \Big)
 + \dots \Big]
\end{align}
where $g_s^{(0)}$ is the string coupling when the dilaton $\Phi=0$, $\nabla_\mu$ denotes the gravitational covariant derivative with respect to the torsion-free connection, the $\dots$ denote higher order terms and $c_i$, $i=1,\dots 9$ are numerical coefficients that can be determined by matching with string scattering amplitudes or
$\sigma$-model conformal invariant conditions~\cite{string,gsw}.  
For constant (or sufficiently slowly moving) dilatons, or weak KR field strengths $\mathcal H_{\mu\nu\rho}$, slowly evolving with the cosmic time,  all of the terms in \eqref{seaPhi} (and those of higher order) are subleading or not contributing to our discussion  Thus we can neglect them and the relevant field equations can be highly simplified in  our case.  Apart from the graviton equation, which we do not write explicitly here, variation of the effective action with respect to both the antisymmetric tensor and dilaton, yields the following field equations:
\begin{align}\label{eqsofmotion1}
{\rm antisymmetric~tensor}:  \qquad  \nabla^\mu \Big(e^{-4\Phi } {\mathcal H}_{\mu\nu\rho} \Big)=0,
\end{align}
\begin{align}\label{eqsofmotion2}
{\rm dilaton}: \frac{2}{\kappa^2}\,  \nabla^\mu\partial_\mu  \Phi - \frac{2}{3}\, e^{-4\Phi}\, {\mathcal H}_{\lambda\mu\nu}{\mathcal H}^{\lambda\mu\nu} + \dfrac{\partial V(\Phi)}{\partial \Phi} = 0~.
\end{align}
In four space-time dimensions, a general solution of  \eqref{eqsofmotion1} is
\begin{align}\label{dualb}
e^{-4\Phi} H_{\mu\nu\rho} \propto \varepsilon_{\mu\nu\rho\sigma} \, \partial^\sigma \overline b(x)\,,
\end{align}
where $\overline b(x)$ is the background of the KR axion field $b(x)$, which was introduced in the main text, as a Lagrange multiplier for the constraint \eqref{modbianchi2}. Indeed, if one ignores higher than quadratic $\mathcal H_{\mu\nu\rho}$ terms in the path integral, after the introduction of the Lagrange multiplier $b$ field~\cite{bmasPRD}, and
considers a saddle point of the action in the (exact) path integration over $\mathcal H_{\mu\nu\rho}$ fields, the  mere use of the equations of motion for the KR field strength yields \eqref{dualb}.

Let us now more specifically address the constant dilaton situation
on which we based our discussion in this work. Upon considering cosmic $b(t)$ fields, with canonically normalised kinetic terms, using \eqref{dualb},
and noting that in our conventions~\cite{bmasPRD} $\frac{2}{3}\, e^{-4\Phi}\,\mathcal H_{\mu\nu\rho} \, \mathcal H^{\mu\nu\rho} = -2(\dot{\overline b})^2 < 0$, we find from \eqref{eqsofmotion2} that $\partial_\mu \Phi \simeq 0$ can be sustained
provided the following relation is reached asymptotically:
\begin{align}\label{phicons}
(\dot{\overline b})^2 = -  \frac12\,\dfrac{\partial V(\Phi)}{\partial\Phi} \Big|_{\Phi\to\Phi_0 \simeq {\rm const}}\gtrsim0\,.
\end{align}
A typical  scenario which could satisfy
\eqref{phicons}, with $\dot b \simeq {\rm constant}$,  hence fulfilling Eq.\, \eqref{lv} of interest to us here, would be, for instance, that of a `run away' pre big bang type~\cite{prebigbang} dilaton potential, in the range where the dilaton slowly approaches a constant value asymptotically and with a decaying trend   $\partial V(\Phi)/\partial\Phi <0$.  Interestingly, this situation characterises N=1 globally supersymmetric theories that can be embedded in a supergravity/superstring framework~\cite{gs}.\footnote{It must be noted that the simplest supercritical strings exponential dilaton potential~\cite{aben}, does not satisfy \eqref{phicons}, but more complicated brane models could~\cite{rizos,dinflation}.} It is then straightforward to see that, under the above conditions, 
the higher order terms in \eqref{seaPhi} have either  vanishing contributions to the equations of motion, or are
subleading, for sufficiently small $\dot b \ll M^2_{\rm Pl}$, as required in the approach of \cite{bmasPRD}, for our cosmic background solutions in a FLRW space time. The upshot is that $\Phi\simeq$const.  appears to be a viable assumption within our framework both as a self-consistent field theory of graviton and KR degrees of freedom or within a generic string inspired approach.


\begin{thebibliography}{99}



\bibitem{bmas}
  S.~Basilakos, N.~E.~Mavromatos and J.~Sol\`a Peracaula,
  Int.\ J.\ Mod.\ Phys.\  D{\bf 28} (2019) 1944002
  ({essay awarded Honorable Mention by the Gravity Research Foundation in the 2019 Essay Competition}).

  \bibitem{bmasPRD}
  S.~Basilakos, N.~E.~Mavromatos and J.~Sol\`a Peracaula,  Phys. Rev.  D{\bf 101} (2020)  045001.


 \bibitem{Planck}
  N.~Aghanim {\it et al.} [Planck Collaboration],
  arXiv:1807.06209;
 { P.A.R.~Ade  {\it et al.} [Planck Collaboration],
Astron. Astrophys. {\bf 594} (2016) A13.}

\bibitem{s8} For recent reviews, see e.g.  L.~Verde, T.~Treu and A.~G.~Riess,
  Nature Astronomy 2019 [arXiv:1907.10625]
 , and references therein;
 J.~Sol\`a Peracaula,
  Int.\ J.\ Mod.\ Phys.\ A {\bf 33} (2018)  1844009.

\bibitem{R19}
 A. G. Riess, {\it et al.}   ApJ, {\bf 876} (2019) 85.

\bibitem{ShapSol} I.~L.~Shapiro and J.~Sol\`a,  JHEP {\bf 0202} (2002) 006;  Phys. Lett. {\bf B 682} (2009) 105.

\bibitem{Fossil07}   J.~Sol\`a,
  J.\ Phys.\ A{\bf 41}  (2008) 164066.


\bibitem{RVMreviews}  J.~Sol\`a,
  J.\ Phys.\ Conf.\ Ser.\  {\bf 453}  (2013) 012015;
  J.~Sol\`a and A. G\'omez-Valent,  Int. J. Mod. Phys. D{\bf 24} (2015) 1541003.



\bibitem{rvmpheno}
J.~Sol\`a, J.~de Cruz P\'erez and A.~G\'omez-Valent,
  EPL {\bf 121}  (2018) 39001;
 MNRAS  {\bf 478}  (2018) 4357;
A.~G\'omez-Valent and  J.~Sol\`a, MNRAS {\bf 478} (2018) 126;
J.~Sol\`a,  A.~G\'omez-Valent and J.~de Cruz P\'erez,
  Phys.\ Lett.\ B {\bf 774}  (2017) 317;  Astrophys. J.  {\bf 836}  (2017) 43;  Astrophys. J. {811} (2015) L14;
  A.~G\'omez-Valent, J. Sol\`a and S. Basilakos, JCAP {\bf 1501} (2015) 004.

\bibitem{JCAP2019}  S.~Basilakos, N.~E.~Mavromatos and J.~Sol\`a Peracaula,  {JCAP {\bf 1912} (2019) 025.}

  \bibitem{BD2019}  	
J.~Sol\`a Peracaula,  A.~G\'omez-Valent,  J.~de Cruz P\'erez and  C.  Moreno-Pulido,
Astrophys. J.  {\bf 886} (2019)  L6.


\bibitem{bamasol} S.~Basilakos, N.~E.~Mavromatos and J.~Sol\`a,
  Universe {\bf 2} (2016) 14
  [arXiv:1505.04434].

\bibitem{rvmInflation}  S.~Basilakos, J.~A.~S.~Lima and J.~Sol\`a,
  MNRAS  {\bf 431}  (2013) 923;  E.~L.~D.~Perico, J.~A.~S.~Lima, S.~Basilakos and J.~Sol\`a,
  Phys.\ Rev.\ D {\bf 88} (2013)  063531.

\bibitem{Yu2019}  J.~Sol\`a Peracaula and H. Yu, 
Gen. Rel. Grav.  52 (2020)  17.

\bibitem{bms}
  M.~de Cesare, N.~E.~Mavromatos and S.~Sarkar,
  Eur.\ Phys.\ J.\ C {\bf 75}  (2015)  514;
  T.~Bossingham, N.~E.~Mavromatos and S.~Sarkar,
  Eur.\ Phys.\ J.\ C {\bf 78}  (2018)  113;
  Eur.\ Phys.\ J.\ C {\bf 79} (2019)  50;
see also: N.~E.~Mavromatos and S.~Sarkar,
  Universe {\bf 5}  (2018)  5
  [arXiv:1812.00504 [hep-ph]].



\bibitem{gsw}
  M.~B.~Green, J.~H.~Schwarz and E.~Witten,
  ``Superstring Theory. Vols. 1: Introduction,''
  Cambridge, Uk: Univ. Pr. ( 1987) 469 P. ( Cambridge Monographs On Mathematical Physics);
``Superstring Theory. Vol. 2: Loop Amplitudes, Anomalies And Phenomenology,''
  Cambridge, Uk: Univ. Pr. ( 1987) 596 P. ( Cambridge Monographs On Mathematical Physics).

\bibitem{string}
  D.~J.~Gross and J.~H.~Sloan,
  Nucl.\ Phys.\ B {\bf 291}  (1987) 41;
R.~R.~Metsaev and A.~A.~Tseytlin,
  Nucl.\ Phys.\ B {\bf 293}  (1987)  385;
M.~C.~Bento and N.~E.~Mavromatos,
  Phys.\ Lett.\ B {\bf 190}  (1987) 105.

\bibitem{kaloper}  M.~J.~Duncan, N.~Kaloper and K.~A.~Olive,
  Nucl.\ Phys.\ B {\bf 387}  (1992)  215.

\bibitem{hehl}
  F.~W.~Hehl, P.~Von Der Heyde, G.~D.~Kerlick and J.~M.~Nester,
  Rev.\ Mod.\ Phys.\  {\bf 48}  (1976) 393.

\bibitem{jackiw}
  R.~Jackiw and S.~Y.~Pi,
  Phys.\ Rev.\ D {\bf 68}  (2003) 104012;
For a more recent review on Chern-Simons modified gravity,  including experimental tests, see:
  S.~Alexander and N.~Yunes,
  Phys.\ Rept.\  {\bf 480}  (2009)  1.

\bibitem{garcia}
  A.~Garcia, F.~W.~Hehl, C.~Heinicke and A.~Macias,
  Class.\ Quant.\ Grav.\  {\bf 21} (2004)  1099.

\bibitem{kanti}
  P.~Kanti, N.~E.~Mavromatos, J.~Rizos, K.~Tamvakis and E.~Winstanley,
  Phys.\ Rev.\ D {\bf 54} (1996) 5049.

\bibitem{stephon}
  S.~H.~S.~Alexander, M.~E.~Peskin and M.~M.~Sheikh-Jabbari,
  Phys.\ Rev.\ Lett.\  {\bf 96}, 081301 (2006);
eConf  C0605151 (2006) 0022
[hep-ph/0701139].




\bibitem{aben}
  I.~Antoniadis, C.~Bachas, J.~R.~Ellis and D.~V.~Nanopoulos,
  Nucl.\ Phys.\ B {\bf 328}, 117 (1989).




\bibitem{lyth}
  D.~H.~Lyth, C.~Quimbay and Y.~Rodriguez,
  JHEP {\bf 0503}  (2005)  016.


\bibitem{paban}  W.~Fischler and S.~Paban,
  JHEP {\bf 0710} (2007) 066.



\bibitem{shapiro}
  G.~de Berredo-Peixoto, L.~Freidel, I.~L.~Shapiro and C.~A.~de Souza,
  JCAP {\bf 1206} (2012)  017.


 \bibitem{adler} S. L. Adler, Phys. \ Rev. \  {\bf 177}, 2246 (1969)
  J. S. Bell and and R. Jackiw, Nuovo Cim. A{\bf 60}, 47 (1969).


\bibitem{qcdaxion} J.~E.~Kim and G.~Carosi,
  Rev.\ Mod.\ Phys.\  {\bf 82}  (2010) 557.



\bibitem{astro} G.~G.~Raffelt,
  Lect.\ Notes Phys.\  {\bf 741} (2008)  51
  [hep-ph/0611350];
A.~Arvanitaki and S.~Dubovsky,
  Phys.\ Rev.\ D {\bf 83}  (2011)  044026;
A.~Arvanitaki, M.~Baryakhtar and X.~Huang,
  Phys.\ Rev.\ D {\bf 91}  (2015) 084011.

\bibitem{latticeqcd} See, for instance:
  G.~Grilli di Cortona, E.~Hardy, J.~Pardo Vega and G.~Villadoro,
  JHEP {\bf 1601}  (2016)  034,
  discussion therein and references thereof.

\bibitem{dine}
  M.~Dine, G.~Festuccia, J.~Kehayias and W.~Wu,
  JHEP {\bf 1101} (2011) 012;
M.~Dine, L.~Stephenson Haskins, L.~Ubaldi and D.~Xu,
  JHEP {\bf 1805} (2018) 171.


\bibitem{svwit}  P.~Svrcek and E.~Witten,
  JHEP {\bf 0606} (2006) 051 and references therein.


  \bibitem{axioninfl} M.~Berg, E.~Pajer and S.~Sjors,
  Phys.\ Rev.\ D {\bf 81} (2010)  103535;
For a review of axion inflation see:
  E.~Pajer and M.~Peloso,
  Class.\ Quant.\ Grav.\  {\bf 30}  (2013) 214002.

\bibitem{arv}
  A.~Arvanitaki, S.~Dimopoulos, S.~Dubovsky, N.~Kaloper and J.~March-Russell,
  Phys.\ Rev.\ D {\bf 81}  (2010)  123530.

\bibitem{sensors} L.~Hui, J.~P.~Ostriker, S.~Tremaine and E.~Witten,
  Phys.\ Rev.\ D {\bf 95}  (2017)  043541;
 Y.V.\, Stadnik and V.V.\, Flambaum,
Phys. Rev. D {\bf 89}  (2014)  043522;
Phys. Rev. Lett. {\bf 115} (2015) 201301;
A.~Aoki and J.~Soda,
  Int.\ J.\ Mod.\ Phys.\ D {\bf 26}  (2016)  1750063;
 C. Abel {\it  et al.},
 Phys. Rev. X {\bf 7} (2017) 041034;
 P.~W.~Graham, D.~E.~Kaplan, J.~Mardon, S.~Rajendran, W.~A.~Terrano, L.~Trahms and T.~Wilkason,
  Phys.\ Rev.\ D {\bf 97} (2018)  055006;
 Z.~Ahmed {\it et al.},
  arXiv:1803.11306 [hep-ex], and references therein.



\bibitem{prebigbang}
  M.~Gasperini and G.~Veneziano,
  Phys.\ Rept.\  {\bf 373} (2003) 1;
%
%
  Astropart.\ Phys.\  {\bf 1} (1993) 317.

\bibitem{gs}
  A.~Font, M.~Klein and F.~Quevedo,
  Nucl.\ Phys.\ B {\bf 605} (2001) 319.

  \bibitem{rizos} N.~E.~Mavromatos and J.~Rizos,
  Int.\ J.\ Mod.\ Phys.\ A {\bf 18} (2003) 57;
 Phys.\ Rev.\ D {\bf 62} (2000) 124004.

\bibitem{dinflation} J.~Ellis, N.~E.~Mavromatos and D.~V.~Nanopoulos,
  Phys.\ Lett.\ B {\bf 732} (2014) 380;
  JCAP {\bf 1411} (2014) 014;
T.~Elghozi, N.~E.~Mavromatos, M.~Sakellariadou and M.~F.~Yusaf,
  JCAP {\bf 1602} (2016) 060.
N.~E.~Mavromatos, M.~Sakellariadou and M.~F.~Yusaf,
  JCAP {\bf 1303} (2013) 015.



  \end{thebibliography}
\end{document}